\documentstyle[aps,prl,preprint,epsf,floats]{revtex}
\tightenlines

\newcommand{\beq}{\begin{equation}}
\newcommand{\eeq}{\end{equation}}
\newcommand{\bqa}{\begin{eqnarray}}
\newcommand{\eqa}{\end{eqnarray}}

\def\square{\vcenter{\vbox{\hrule height.4pt
          \hbox{\vrule width.4pt height8pt
          \kern8pt\vrule width.4pt}\hrule height.4pt}}}

\def\sumint{\hbox{$\sum$}\!\!\!\!\!\!\int}

\begin{document}

\preprint{
\vbox{\halign{&##\hfil\cr
        & hep-ph/0007159 \cr
&\today\cr }}}

\title{Screened Perturbation Theory to Three Loops}

\author{Jens O. Andersen and Eric Braaten}
\address{Physics Department, Ohio State University, Columbus OH 43210, USA}

\author{Michael Strickland}
\address{Physics Department, University of Washington, Seattle  WA 98195-1560}
\maketitle

\begin{abstract}
{\footnotesize 
The thermal physics 
of a massless scalar field with a $\phi^4$ 
interaction is studied
within screened perturbation theory (SPT).
In this method the perturbative expansion is reorganized by adding and 
subtracting a mass term in the lagrangian.
We consider several different mass prescriptions that generalize the 
one-loop gap equation to two-loop order.
We calculate the pressure and entropy
to three-loop order and the screening mass to two-loop order.
In contrast to the weak-coupling expansion, 
the SPT-improved approximations 
appear to converge even for rather large values of the coupling constant.
}
\end{abstract}

\newpage
\section{Introduction}
If we have a weakly-coupled quantum field theory in equilibrium at temperature
$T$, we should be able to use perturbation theory as a
quantitative tool to study its properties.
In the case of a massless theory with coupling constant $g$, 
the naive perturbative expansion in powers of $g^2$ breaks down because
of collective effects such as screening. 
However, the perturbative expansion can be reorganized 
into a weak-coupling expansion in powers of $g$ either by using 
resummation methods or alternatively by using effective field theory.
It is reasonable to assume
that this weak-coupling expansion provides a useful
asymptotic expansion for sufficiently small values of $g$.

Only in recent years has the calculational technology of thermal quantum
field theory advanced to the point where this assumption can be tested.
Unfortunately, the assumption seems to be false.
One would expect the thermodynamic functions, such as the pressure, to be 
among the quantities with the best-behaved weak-coupling expansion, since
collective effects are suppressed by several powers of $g$.
However, in recent years, the thermodynamic functions have been calculated to
order $g^5$ for massless scalar theories
\cite{arnold-zhai,Parwani-Singh,Braaten-Nieto:scalar},
abelian gauge theories \cite{Parwani,Andersen},
and nonabelian gauge theories 
\cite{arnold-zhai,Kastening-Zhai,Braaten-Nieto:QCD}.  
The weak-coupling expansions show no sign of converging
even for extremely small values of $g$. 
There is already a hint of the
problem in the $g^3$ correction, which has the opposite 
sign and is relatively large compared to the $g^2$ coefficient.
The large size of the $g^3$ contribution is not necessarily fatal, 
since it is the first term that takes into account collective effects.
An optimist might still hope that higher-order corrections would 
be well-behaved. This optimism has been dashed by the explicit calculation
of the $g^4$ and $g^5$ terms.

\begin{figure}[htb]
\epsfysize=8cm
\centerline{\epsffile{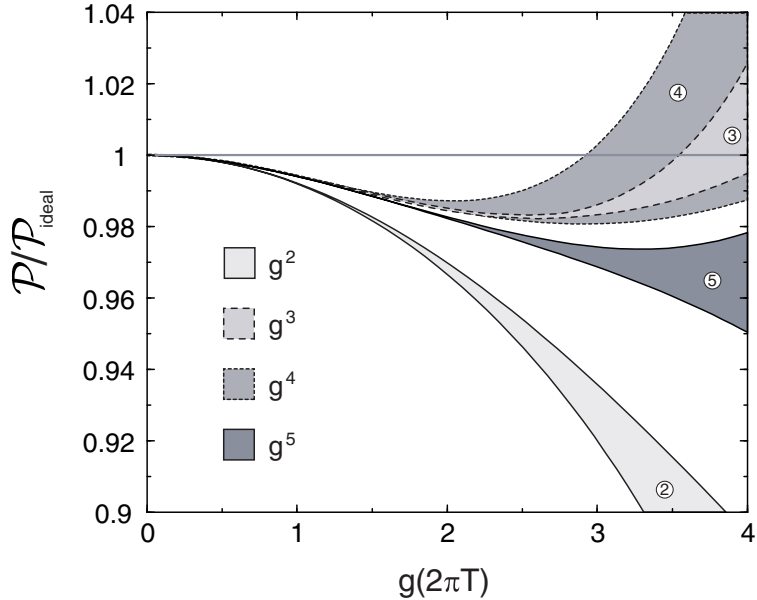}}
\vspace{3mm}
\caption[a]{Weak-coupling expansion to orders $g^2$, $g^3$, $g^4$, and $g^5$
for the pressure normalized to that of an ideal gas as a function
of $g(2\pi T)$.}
\label{fpert}
\end{figure}

For a massless scalar field theory with a $g^2\phi^4/4!$ interaction, 
the weak-coupling expansion for the pressure to 
order $g^5$ is~\cite{arnold-zhai,Parwani-Singh,Braaten-Nieto:scalar}
\bqa\nonumber
{\cal P} &=& {\cal P}_{\rm ideal} \left[
1-{5\over4}\alpha+{5\sqrt{6}\over3}\alpha^{3/2}+{15\over4}
\left(\log{\mu\over2\pi T}+0.40\right)\alpha^2\right.\\ 
&&\left.-{15\sqrt{6}\over2}\left(\log{\mu\over2\pi T}-{2\over3}\log\alpha
-0.72\right)\alpha^{5/2}+{\cal O}(\alpha^3\log\alpha)\right]\;,
\eqa
where ${\cal P}_{\rm ideal} = (\pi^2/90)T^4$
is the pressure of an ideal gas of free massless bosons,
$\alpha=g^2(\mu)/16\pi^2$, and $g(\mu)$ is the 
$\overline{\rm MS}$ coupling constant at the renormalization scale $\mu$.
In Fig.~\ref{fpert}, we show the successive perturbative approximations to
${\cal P}/{\cal P}_{\rm ideal}$ as a function of $g(2\pi T)$. Each partial
sum is shown as a band obtained by varying $\mu$ 
from $\pi T$ to $4\pi T$.
To express $g(\mu)$ in terms of
$g(2\pi T)$, we use the numerical 
solution to the renormalization group equation
$\mu{\partial\over \partial\mu}\alpha=\beta(\alpha)$ 
with a five-loop beta  function~\cite{Kleinert}:
\bqa
\label{rg2}
\mu{\partial\over \partial\mu}\alpha=
3\alpha^2-{17\over3}\alpha^3+32.54\alpha^4-271.6\alpha^5+
2848.6\alpha^6\;.
\eqa
The lack of convergence of the weak-coupling expansion 
is evident in Fig.~\ref{fpert}.
The band obtained by varying $\mu$ by a factor of two is not 
necessarily a good measure of the error, 
but it is certainly a lower bound on the theoretical error.
Another indicator of the  theoretical error is the deviation 
between successive approximations.  We can infer from Fig.~\ref{fpert} 
that the error grows rapidly when $g(2 \pi T)$ exceeds 1.5.

\begin{figure}[htb]
\epsfysize=8cm
\centerline{\epsffile{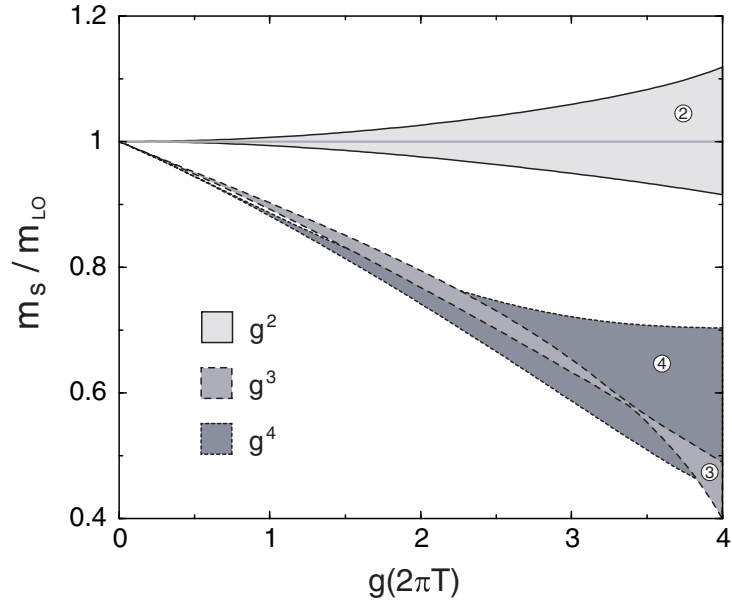}}
\vspace{3mm}
\caption[a]{Weak-coupling expansion to orders $g^2$, $g^3$, and $g^4$
for the screening mass normalized to the leading-order expression as a
function of $g(2\pi T)$.}
\label{mspert}
\end{figure}

A similar behavior can be seen in the weak-coupling expansion 
for the screening mass, which has been calculated to 
next-to-next-to-leading
order in $g$~\cite{Braaten-Nieto:scalar}:
\bqa
\label{mpert}
m_s^2={2\pi^2\over3}\alpha T^2\Bigg\{1-\sqrt{6}\alpha^{1/2}
- \left[ 3 \log{\mu\over2\pi T}- 2 \log\alpha - 6.4341 \right] \alpha
+{\cal O}(\alpha^{3/2})\Bigg\}\;.
\eqa
In Fig.~\ref{mspert}, we show the screening mass $m_s$ 
normalized to the leading order result $m_{LO}=g(2\pi T)T/\sqrt{24}$
as a function of $g(2\pi T)$, for each of the three
successive approximations to $m_s^2$.
The bands correspond to varying $\mu$
from $\pi T$ to $4\pi T$. The poor convergence is again evident.
The pattern is similar to that in Fig.~\ref{fpert}, 
with a large deviation between the order-$g^2$ and order-$g^3$ 
approximations and a large increase in the size of the band for $g^4$.

There are many possibilities for reorganizing the weak-coupling expansion 
to improve
its convergence. One possibility is to use Pad\'e 
approximants~\cite{Pade}. 
This method is limited 
to observables like the pressure,
for which several terms in the weak-coupling expansion are known.
Its application is further complicated by the appearance of logarithms
of the coupling constant in the coefficients of the weak-coupling expansion.
However, the greatest problem with Pad\'e approximants is that, with
no understanding of the analytic behavior of ${\cal P}$ at strong coupling,
it is little more than a numerological recipe.

An alternative with greater physical motivation 
is a self-consistent approach \cite{Baym}.
Perturbation theory can be reorganized by expressing the free energy 
as a stationary point of a functional $\Omega$ 
of the exact self-energy function $\Pi(p_0,{\bf p})$ 
called the thermodynamic potential \cite{Luttinger-Ward}.
Since the exact self-energy is not known, $\Pi$ can be regarded as a 
variational function. 
The ``$\Phi$-derivable'' prescription of Baym~\cite{Baym}
is to truncate the perturbative expansion 
for the thermodynamic potential $\Omega$ and to determine
$\Pi$ self-consistently as a stationary point of $\Omega$. This gives an
integral equation for $\Pi$ which is difficult to solve numerically,
except in cases where $\Pi$ is momentum independent. 
In relativistic field theories, there are additional
complications from ultraviolet
divergences. A more tractable approach is to find an approximate solution
to the integral equations that is accurate only in the weak-coupling limit.
Such an approach has been applied by Blaizot, Iancu, and
Rebhan to massless scalar field theories and gauge 
theories~\cite{comp1,BIR}.

Another approach that is also variational in spirit is
{\it screened perturbation theory} (SPT) introduced by Karsch, Patk\'os
and Petreczky~\cite{K-P-P}. This approach is less ambitious than the 
$\Phi$-derivable approach.
Instead of introducing a variational function, it introduces a single
variational parameter $m$. This parameter has a simple and obvious
physical interpretation as a thermal mass. 
The advantage of screened perturbation theory is that it is very easy to apply.
Higher order corrections are tractable, so one can test whether it improves
the convergence of the weak-coupling expansion.
Karsch, Patk\'os and Petreczky applied screened perturbation theory
to a massless scalar field theory with a $\phi^4$ interaction,
computing the two-loop
pressure and the three-loop pressure in the large-$N$ limit.
In both cases, they used a one-loop gap equation as their prescription for the
mass. 
Their three-loop calculation was not a very stringent test of the method, 
because the large-$N$ limit
suppresses self-energy diagrams that depend on the momentum.

In this paper, we present a thorough study of screened perturbation theory
for a massless scalar field theory with a $\phi^4$ interaction.
We calculate the pressure and entropy to three loops and the screening mass to 
two loops using SPT.
We consider several generalizations of the one-loop gap equation to
two loops. 
Inserting the solutions to the gap equations for $m$
into the SPT expansions,         
we obtain the SPT-improved approximations to the pressure,
the screening mass, and the entropy.

The paper is organized as follows. In section~\ref{systematics}, 
we describe the systematics of screened perturbation theory. 
In section~\ref{mpre}, we discuss the possible prescriptions that can be used
to generalize the one-loop gap equation to higher orders.
We calculate the free energy to three-loop order in section~\ref{1} 
and the screening mass to two-loop order in section~\ref{masscal}.
In section~\ref{gapeqs}, we study three generalizations of the
one-loop gap equation to two-loop order.
In section~\ref{observables}, we study the convergence of the SPT-improved
results for the pressure, screening mass, and entropy.
In section~\ref{conclusions}, we summarize and conclude.
We have collected the necessary sum-integrals in an appendix. 

\section{Screened Perturbation Theory}
\label{systematics}

The lagrangian density for a massless scalar field with a $\phi^4$
interaction is
\bqa
\label{l1}
{\cal L}={1\over2}\partial_{\mu}\phi\partial^{\mu}\phi
-{1\over 24}g^2\phi^4+\Delta{\cal L}\;,
\label{barel}
\eqa
where $g$ is the coupling constant and $\Delta{\cal L}$ includes counterterms.
The conventional perturbative expansion in powers of $g^2$ generates
ultraviolet divergences, and the counterterm $\Delta{\cal L}$
must be adjusted to cancel the divergences order by order in $g^2$.
If we use dimensional regularization in $d=3-2\epsilon$ spatial dimensions
and minimal subtraction to remove the ultraviolet divergences, the 
counterterms have the form
\bqa
\label{counter}
\Delta{\cal L}={1\over2}\left(Z_{\phi}-1\right)\partial_{\mu}\phi
\partial^{\mu}\phi-{1\over24}\Delta g^2\phi^4,
\eqa
where $\Delta g^2=\left(Z_{\phi}^2Z_g-1\right)g^2$, and
$Z_{\phi}$ and $Z_{g}$ are power series in $g^2$ whose coefficients have poles
in $\epsilon$.
At nonzero temperature, the conventional perturbative expansion also generates
infrared divergences. They can be removed by resumming the higher order 
diagrams that generate a thermal mass of order $gT$ for the scalar particle.
This resummation changes the perturbative series from an expansion in powers
of $g^2$ to an expansion in powers of $\left(g^2\right)^{1/2}=g$.

Screened perturbation theory, which was introduced by Karsch, Patk\'os
and Petreczky~\cite{K-P-P}, is simply a reorganization of the perturbation
series for thermal field theory.
It can be made more systematic by using a framework called 
``optimized perturbation theory'' that Chiku and Hatsuda~\cite{Chiku-Hatsuda}
have applied to a spontaneously broken scalar field theory. The lagrangian
density is written as
\bqa
\label{SPT}
{\cal L}_{\rm SPT}=-{\cal E}_0 + {1 \over 2} \partial_\mu\phi\partial^\mu\phi
	-{1\over2}(m^2-m_1^2)\phi^2 - {1\over24} g^2 \phi^4
	+\Delta{\cal L}
	+\Delta{\cal L}_{\rm SPT}\;,
\eqa
where ${\cal E}_0$ is a vacuum energy density
parameter and we have added and subtracted
mass terms.
If we set ${\cal E}_0=0$ and $m_1^2=m^2$, we recover the original 
lagrangian~(\ref{barel}).
Screened perturbation theory is defined by taking $m^2$ to be of order 
$g^0$ and $m_1^2$ to be of order $g^2$, expanding systematically in powers
of $g^2$, and setting $m_1^2=m^2$ at the end of the calculation.
This defines a reorganization of perturbation theory in which the expansion
is around the free field theory defined by
\bqa
\label{freesca}
{\cal L}_{\rm free}=-{\cal E}_0+{1\over2}\partial_{\mu}\phi\partial^{\mu}\phi
-{1\over2}m^2\phi^2\;.
\eqa
The interaction term is
\bqa
\label{scaint}
{\cal L}_{\rm int}= {-}{1\over24} g^2 \phi^4 + {1\over2}m_1^2\phi^2
+\Delta{\cal L}+\Delta{\cal L}_{\rm SPT}\;.
\eqa
At each order in screened perturbation theory,
the effects of the $m^2$ term in~(\ref{freesca}) are included to all orders.
However, when we set $m_1^2 = m^2$, the dependence on $m$ 
is systematically subtracted out at higher
orders in perturbation theory by the $m_1^2$ term in~(\ref{scaint}).
At nonzero temperature, screened perturbation theory does not generate
any infrared divergences, because the mass parameter $m^2$ in the
free lagrangian~(\ref{freesca}) provides an infrared cutoff. 
The resulting perturbative expansion is therefore a power series in $g^2$
and $m_1^2=m^2$ whose coefficients depend on the mass parameter $m$.

This reorganization of perturbation theory generates new ultraviolet
divergences, but they can be canceled by the additional counterterms
in $\Delta{\cal L}_{\rm SPT}$. The renormalizability of the lagrangian 
in~(\ref{SPT}) guarantees that the only counterterms required
are proportional to 1, $\phi^2$, $\partial_{\mu}\phi\partial^{\mu}\phi$,
and $\phi^4$.
With dimensional regularization and minimal subtraction, the coefficients
of these operators are polynomials in $\alpha=g^2/16\pi^2$ and $m^2-m_1^2$.
The extra counterterms required to remove the additional
ultraviolet divergences are 
\bqa
\Delta{\cal L}_{\rm SPT}=-\Delta{\cal E}_0-{1\over2}\left(
\Delta m^2-\Delta m_1^2
\right)\phi^2\;.
\eqa
The vacuum energy counterterm has the form 
$\Delta{\cal E}_0=Z_{E}\left(m^2-m_1^2\right)^2$, where $Z_{E}$ is a power
series in $\alpha$ whose coefficients have poles in
$\epsilon$. The mass counterterms have the form 
$\Delta m^2=(Z_{\phi}Z_m-1)m^2$ and
$\Delta m^2_1=(Z_{\phi}Z_m-1)m^2_1$, where 
$Z_{\phi}$ is the same wavefunction renormalization constant that appears
in~(\ref{counter}) and
$Z_m$ is also a power series in $\alpha$ whose coefficients have poles in
$\epsilon$.

Several terms in the power series expansions of the
counterterms are known from previous calculations at zero temperature.
The counterterms $\Delta g^2$ and $\Delta m^2$ are known to order 
$\alpha^5$~\cite{Kleinert}. 
We will need the coupling constant counterterm only to leading
order in $\alpha$:
\bqa
\Delta g^2=\left[
{3\over2\epsilon}\alpha+...
\right]g^2\;.
\eqa
We need the mass counterterms $\Delta m^2$ and $\Delta m^2_1$
to next-to-leading order and 
leading order in $\alpha$, respectively:
\bqa
\label{dmm}
\Delta m^2&=&\left[{1\over2\epsilon}\alpha+\left({1\over2\epsilon^2}
-{5\over24\epsilon}\right)\alpha^2+...
\right]m^2\;, \\
\label{d1m1}
\Delta m^2_1&=&\left[{1\over2\epsilon}\alpha+...
\right]m^2_1\;.
\eqa
The counterterm for $\Delta{\cal E}_0$
has been calculated to order $\alpha^4$~\cite{kastening}.
We will need its expansion only to second order 
in $\alpha$ and $m_1^2$:
\bqa\nonumber
(4\pi)^2\Delta{\cal E}_0&=&\left[
{1\over4\epsilon}+{1\over8\epsilon^2}\alpha
+\left(
{5\over48\epsilon^3}-{5\over72\epsilon^2}+{1\over96\epsilon}
\right)\alpha^2
\right]m^4
\\
&&
\label{de}
-2\left[{1\over4\epsilon}+{1\over8\epsilon^2}\alpha\right]m_1^2m^2
+{1\over4\epsilon}m_1^4\;.
\eqa

\section{Mass Prescriptions}
\label{mpre}

The mass parameter $m$ in screened perturbation theory is completely arbitrary.
To complete a calculation in screened perturbation theory, it is necessary
to specify $m$ as a function of $g$ and $T$. 
One of the complications from the ultraviolet divergences is that
the parameters ${\cal E}_0$, $m^2$, $g^2$, and $m_1^2$ all become
running parameters that depend on a renormalization scale $\mu$.
In our prescription for recovering the original theory, we must therefore
specify the renormalization scale $\mu_*$
at which the lagrangian~(\ref{SPT}) reduces to~(\ref{l1}).
The prescription can
be written
\bqa
{\cal E}_0(\mu_*)&=0& \, , \\ 
m^2(\mu_*)&=&m_1^2(\mu_*)=m^2_{*}(T),
\eqa
where $m_{*}(T)$ is some prescribed function of the temperature.
This is the only point where temperature enters into SPT. 
We proceed to discuss the possible prescriptions for
$m_{*}(T)$.

The prescription of
Karsch, Patk\'os, and Petreczky for $m_*(T)$ is the solution to the
one-loop gap equation:
\bqa
\label{pet}
m_{*}^2={1\over2}\alpha(\mu_*)\left[
J_1(m_*/T)T^2-\left(2\log{\mu_*\over m_*}+1\right)m_*^2
\right]
\;,
\eqa
where the function $J_1(x)$ is defined in~(\ref{jn2}).
Their choice for the scale was $\mu_*=T$.
In the weak-coupling limit, the solution to~(\ref{pet}) is 
$m_*=g(\mu_*)T/\sqrt{24}$. 

There are many possibilities for generalizing~(\ref{pet}) to higher orders
in $g$. One class of possibilities 
is to identify $m_*$
with some physical mass in the system.  The simplest choice is the 
{\it screening mass} $m_s$ defined by the location of the pole in the static
propagator:
\bqa
\label{scrdef}
{\bf p}^2 + m^2+\Pi(0,{\bf p}) = 0 \hspace{1cm} 
	{\rm at} \;\;\; {\bf p}^2=-m_s^2 \;,
\eqa
where $\Pi(p_0,{\bf p})$ is the self-energy function.
Another choice is the rest mass of the quasiparticle:
$m_q={\rm Re}\;\omega(0)$, where
$\omega(p)$ is the quasiparticle dispersion relation which satisfies 
$-\omega^2+p^2+\Pi(i(\omega+i\epsilon),{\bf p})=0$. 
The quasiparticle mass is more difficult to calculate than the screening mass.

Another mass prescription that generalizes~(\ref{pet}) to higher orders is to
identify $m_*$ with the {\it tadpole mass} defined by
$m_t^2=g^2\langle\phi^2\rangle$.
This can also be expressed as
a derivative of the free energy:
\bqa
\label{diffdef}
m_t^2={\!\!\!\!\partial\over \partial m^2}{\cal F}(T,g,m,m_1,\mu)\Bigg|_{m_1=m}\;,
\eqa
where the partial derivative is taken before setting $m_1=m$. 
An advantage of the tadpole mass is that $\langle\phi^2\rangle$ is easier
to calculate at higher orders than the self-energy $\Pi$.

There is another class of prescriptions that is variational in spirit.
The results of SPT would be independent of $m$
if they were calculated to all orders. This suggests choosing $m$ to
minimize the dependence of some physical quantity on $m$.
Taking that physical quantity to be the free energy, the prescription is
\bqa
\label{vmass}
{\!\!\!d\over dm^2}{\cal F}(T,g(\mu),m,m_1=m,\mu)=0\;.
\eqa
We will refer to the solution $m_v$ to this equation as the 
{\it variational mass}.

One mass prescription that may seem appealing is to choose
$m_*(T)$ so that the perturbative approximation 
is thermodynamically consistent~\cite{goren}.
Given a diagrammatic expansion for ${\cal F}$, 
the entropy density ${\cal S}$ has a diagrammatic expansion given 
by
\bqa
\label{e}
{\cal S}_{\rm diag}&=&-{\!\!\partial\over \partial T}{\cal F}
(T,g,m,m_1,\mu)
\;,
\eqa
where the partial derivative $\partial/\partial T$ is taken with all
the other variables $g$, $m$, $m_1$, and $\mu$ held fixed. 
The entropy density can also be defined by the thermodynamic relation
\bqa
\label{s}
{\cal S}_{\rm thermo}&=&-{\!\!d\over dT}{\cal F}(T,g(\mu),m=m_*,m_1=m_*,\mu)\;. 
\eqa
The total derivative takes into account the explicit dependence on $T$,
the $T$-dependence of $m_*(T)$, and also the $T$-dependence of the running
coupling constant if we choose a scale $\mu$ that depends on $T$.
If the thermodynamic expansions for ${\cal F}$ and ${\cal S}$ were known to 
all orders, there would be no dependence on $m$ or $\mu$, and~(\ref{e}) 
and~(\ref{s}) would be equivalent.
If the diagrammatic expansion is truncated and if any of the parameters
$g$, $m$, $m_1$, and $\mu$ is allowed to depend on $T$, then 
${\cal S}$ may not satisfy~(\ref{s}). 
An approximation is called {\it thermodynamically
consistent} if ${\cal S}$ satisfies~(\ref{s}) exactly.
This requires
\bqa
\label{consi}
{dg\over dT}{\partial{\cal F}\over\partial g}+
{d\mu\over dT}{\partial{\cal F}\over\partial\mu}+
{dm\over dT}{\partial{\cal F}\over\partial m}+
{dm_1\over dT}{\partial{\cal F}\over\partial m_1}=0\;.
\eqa
If ${\cal F}$ were known to all orders, it would be independent of 
$m$ and $m_1$ at $m=m_1$.
Thermodynamic consistency could then be guaranteed by taking 
the scale $\mu$ to be any function of $T$ and choosing 
$g(\mu)$ to be the running coupling constant at that scale.
If we only have a perturbative approximation to 
${\cal F}$,~(\ref{consi}) is satisfied only up to higher order corrections.
One way
to guarantee thermodynamic consistency 
is to choose $\mu=am$ with $a$ a constant and impose the
condition
\bqa
{\!\!\!d\over dm^2}{\cal F}(T,g(am),m,m_1=m,\mu=am)=0\;.
\eqa
This differs from the variational gap equation~(\ref{vmass})
only in that we have set $\mu=am$ before differentiating.
This equation does not reduce to the one-loop gap equation~(\ref{pet}) at
leading order, so we will not consider it any further. We will be satisfied
by approximations that are thermodynamically consistent only up to higher
orders in perturbation theory.

\section{Free Energy to Three Loops}\label{1}

\begin{figure}[htb]
\epsfysize=3.0cm
\centerline{\epsffile{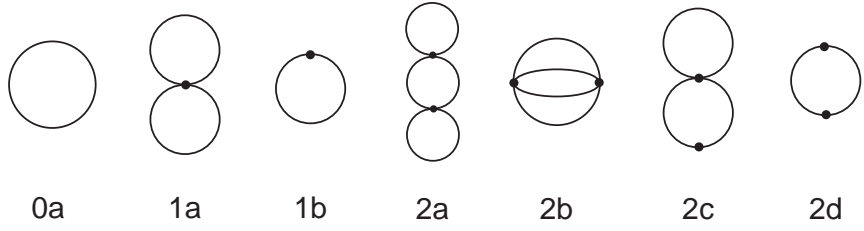}}
\vspace{3mm}
\caption[a]{Diagrams for the one-loop (0a), two-loop (1a and 1b), and
three-loop (2a, 2b, 2c and 2d) free energy.}
\label{2ldiagrams}
\end{figure}

In this section, we calculate the pressure and entropy density
to three loops in screened perturbation theory.  
The diagrams for the free energy that are included at this order 
are those shown in Fig.~\ref{2ldiagrams} 
together with diagrams involving counterterms.

\subsection{One-loop free energy}
\label{1.1}

The free energy at leading order in $g^2$ is
\bqa
\label{freebos}
{\cal F}_0 &=& {\cal E}_0 + {\cal F}_{\rm 0a} 
+ \Delta_0{\cal E}_0 \; , 
\eqa
where $\Delta_0{\cal E}_0$ is the term of order $g^0$ in the 
vacuum energy counterterm~(\ref{de}). 
The expression for diagram 0a in Fig.~\ref{2ldiagrams} is
\bqa
\label{0a}
{\cal F}_{\rm 0a} &=& {1\over2}\sumint_P\log\left[P^2+m^2\right]\;.
\eqa
The sum-integral in~(\ref{0a}) 
is over the Euclidean momentum $P=(\omega_n,{\bf p})$ and we define
$P^2 = {\bf p}^2 + \omega_n^2$. The sum-integral includes a sum over
Matsubara frequencies $\omega_n = 2 \pi n T$
and a dimensionally regularized integral over the momentum ${\bf p}$
with a measure that is defined in Appendix \ref{sumintegrals}.  
In dimensional regularization with $3-2\epsilon$ spatial dimensions,
the diagrams for ${\cal F}$ have dimensions (energy)$^{4-2\epsilon}$.
To obtain the renormalized free energy density with dimensions (energy)$^4$,
we multiply the diagrams by $\mu^{2\epsilon}$, where $\mu$ is an arbitrary
renormalization scale, before taking the limit $\epsilon\to 0$.
The coupling constant in dimensional regularization is $g\mu^{\epsilon}$,
where $g$ is the dimensionless renormalized coupling constant. 
Including the overall
factor of $\mu^{2\epsilon}$ and the factor of $\mu^{\epsilon}$
from the coupling constants, there is a factor of $\mu^{2\epsilon}$
for each sum-integral. We choose to absorb this factor into the measure of the
sum-integral.

The sum-integral in (\ref{0a})  is expressed as a function of $\epsilon$
in the Appendix. It has a pole at $\epsilon=0$. 
The result for the diagram is 
\bqa
\label{res0a}
{\cal F}_{\rm 0a} &=&
-{1\over4(4\pi)^2}\left({\mu\over m}\right)^{2\epsilon}
\Bigg\{\left[{1\over\epsilon}+{3\over2}
+
{21+\pi^2\over12}
\epsilon
+
{45+3\pi^2+4\psi^{\prime\prime}(1)\over24}
\epsilon^2
\right]m^4+2J_0T^4
\Bigg\}\;,
\eqa
where $J_0$ is the function of $m/T$ defined in (\ref{jndef}).
We have
kept all terms that 
contribute through order $\epsilon^2$, 
because they enter into higher order diagrams involving counterterms.
The pole in $\epsilon$ in~(\ref{res0a}) is canceled by the 
zeroth order term $\Delta_0{\cal E}_0$
in the counterterm~(\ref{de}).
The final result for the one-loop free energy is
\bqa
\label{f1loop}
(4\pi)^2{\cal F}_0&=&(4\pi)^2{\cal E}_0-{1\over8}(2L+3)m^4
-{1\over2}J_0T^4\;,
\eqa
where $L=\log(\mu^2/m^2)$ and $J_0$ can now be replaced by its value at 
$\epsilon=0$, which is given in~(\ref{jn2}).

\subsection{Two-loop free energy}
\label{1.2}

\noindent
The contribution to the free energy of order
$g^2$ is
\bqa
\label{twobare}
{\cal F}_1 &=& {\cal F}_{\rm 1a} + 
{\cal F}_{\rm 1b}+ \Delta_1{\cal E}_0 +
{\partial {\cal F}_{\rm 0a} \over \partial m^2} \Delta_1 m^2\;,
\label{f2}
\eqa
where $\Delta_1{\cal E}_0$ and $\Delta_1 m^2$ are the terms of order $g^2$ 
in the counterterms~(\ref{dmm}) and~(\ref{de}), respectively.
The expressions for the diagrams 1a and 1b in Fig.~\ref{2ldiagrams}
are
\bqa
\label{two}
{\cal F}_{\rm 1a}&=&{1\over8}g^2\left(\sumint_P{1\over P^2+m^2}\right)^2\,,\\
\label{count1}
{\cal F}_{\rm 1b}&=&-{1\over 2}m_1^2\sumint_P{1\over P^2+m^2}\;.
\eqa
The results for the diagrams can be expressed as 
\bqa\nonumber
{\cal F}_{\rm 1a} &=&{\alpha\over8(4\pi)^2}
\left({\mu\over m}\right)^{4\epsilon}\Bigg\{\left[{1\over\epsilon^2}
	+{2\over\epsilon}
+{18+\pi^2\over6}+{12+\pi^2+\psi^{\prime\prime}(1)\over3}
\epsilon\right]m^4
\label{f1alt}\\
&&\hspace{3.04cm}
-2\left[{1\over\epsilon}+1+{12+\pi^2\over12}\epsilon\right]J_1m^2T^2+J_1^2T^4
\Bigg\} 
\;,\\
\label{f1blt}
{\cal F}_{\rm 1b} &=& 
-{m_1^2\over2(4\pi)^2}\left({\mu\over m}\right)^{2\epsilon}
\Bigg\{-\left[{1\over\epsilon}+1+{12+\pi^2\over12}
\epsilon\right]m^2+J_1T^2\Bigg\}\;,
\eqa
where $\alpha=g^2/16\pi^2$.
We have kept all terms that contribute through order $\epsilon$, because
they are needed for counterterm diagrams in the three-loop free energy.
The poles in $\epsilon$ in~(\ref{f1alt}) and~(\ref{f1blt})
are canceled by the counterterms 
in~(\ref{twobare}).
The final result for the two-loop free energy is
\bqa
(4\pi)^2{\cal F}_1&=&
{1\over2}\left[(L+1)m^2-J_1T^2\right]m_1^2 
+{1\over8}\alpha\left[
(L+1)m^2-J_1T^2
\right]^2\;.
\label{f2loop}
\eqa

\subsection{Three-loop free energy}
\label{3}

The contribution to the free energy of order $g^4$ is
\bqa
{\cal F}_2 &=& {\cal F}_{\rm 2a} + 
{\cal F}_{\rm 2b} +{\cal F}_{\rm 2c} + {\cal F}_{\rm 2d}
+ \Delta_2{\cal E}_0 +
{\partial {\cal F}_{\rm 0a} \over \partial m^2} \Delta_2 m^2 
+{1\over2}
{\partial^2{\cal F}_{\rm 0a}\over (\partial m^2)^2} 
\left(\Delta_1 m^2\right)^2 
\nonumber \\
&& 
\hspace{5mm}
\label{3unren}
+
\left({\,\partial {\cal F}_{\rm 1a} \over \partial m^2}
+{\,\partial {\cal F}_{\rm 1b} \over \partial m^2}
\right) \Delta_1 m^2 +
{{\cal F}_{\rm 1a} \over g^2} \Delta_1 g^2
+{{\cal F}_{\rm 1b}\over m_1^2}\Delta_1m_1^2
\;,
\label{f3}
\eqa
where we have included all the appropriate counterterms. 
The expressions for the diagrams 2a, 2b, 2c, and 2d in 
Fig.~\ref{2ldiagrams} are
\bqa
{\cal F}_{\rm 2a}&=&-{1\over16}g^4
	\left(\sumint_{P}{1\over P^2+m^2}\right)^2 
	\sumint_{Q}{1\over(Q^2+m^2)^2} \; , \\
{\cal F}_{\rm 2b}&=&-{1\over48}g^4
	\sumint_{PQR}{1\over(P^2+m^2)(Q^2+m^2)(R^2+m^2)((P+Q+R)^2+m^2)} \;, \\ 
{\cal F}_{\rm 2c}&=&{1\over4}g^2m_1^2
	\sumint_{P}{1\over P^2+m^2}
	\sumint_{Q}{1\over(Q^2+m^2)^2} \; , \\
{\cal F}_{\rm 2d}&=&-{1\over4}m_1^4
	\sumint_{P}{1\over(P^2+m^2)^2} \; .
\eqa
The results for these diagrams in the limit $\epsilon\rightarrow 0$ are
\bqa\nonumber
{\cal F}_{\rm 2a} &=& 
-{\alpha^2\over16(4\pi)^2} 
	\left({\mu\over m}\right)^{6\epsilon}
\left\{\left[{1\over\epsilon^3}+{2\over\epsilon^2}
+{12+\pi^2\over4\epsilon}+{8+\pi^2+
\psi^{\prime\prime}(1)\over2}\right]m^4\right.\\ \nonumber
&&\hspace{3.45cm}
+\left[{1\over\epsilon^2}+{2\over\epsilon}+{18+\pi^2\over6}\right]J_2m^4
-2\left[{1\over\epsilon^2}+{1\over\epsilon}+{6+\pi^2\over6}\right]J_1m^2T^2 \\
&&
\left.\hspace{3.45cm}
-2\left[{1\over\epsilon}+1\right]J_1J_2m^2T^2
+{1\over\epsilon}J_1^2T^4
+J_1^2J_2T^4\right\}\;,\\ \nonumber
{\cal F}_{\rm 2b} &=& 
-{\alpha^2\over48(4\pi)^2}
\left({\mu\over m}\right)^{6\epsilon}
\left\{\left[{2\over\epsilon^3}+{23\over3\epsilon^2}+{35+\pi^2\over2\epsilon}
+C_0\right]m^4-
\left[{6\over\epsilon^2}+{17\over\epsilon}-4C_1\right]J_1m^2T^2\right.\\
&&\hspace{3.45cm}
\left.+\left[{6\over\epsilon}+12\right]J_1^2T^4
+\left[6K_2+4K_3\right]T^4\;
\right\},
\\ 
{\cal F}_{\rm 2c} &=& 
{\alpha m_1^2\over4(4\pi)^2} 
	\left({\mu\over m}\right)^{4\epsilon}
\left\{
-\left[{1\over\epsilon^2}+{1\over\epsilon}+{6+\pi^2\over6}\right]m^2
-\left({1\over\epsilon}+1\right)J_2m^2
+{1\over\epsilon}J_1T^2+J_1J_2T^2
\right\}\;,\\ 
{\cal F}_{\rm 2d} &=& 
-{m_1^4\over4(4\pi)^2}
	\left({\mu\over m}\right)^{2\epsilon}
\left\{{1\over\epsilon}+J_2\right\}\;,
\eqa
The poles in $\epsilon$ are canceled by the counterterms in~(\ref{3unren}).
The final result for the free energy is
\bqa\nonumber
(4\pi)^2{\cal F}_2&=&
-{1\over4}\left(L+J_2\right)m^4_1
-{\alpha\over4}
(L+J_2)\left[
(L+1)m^2-J_1T^2
\right]m_1^2\\ \nonumber
&&-
{1\over48}\alpha^2
\left[\left(
5L^3+17L^2+\mbox{$41\over2$}L-23-\mbox{$23\over12$}\pi^2
	-\psi^{\prime\prime}(1)+C_0+3(L+1)^2J_2
\right)m^4\right.
\\ \nonumber
&&
\left.
\hspace{1cm}-\left(12L^2+28L-12-\pi^2-4C_1
+6(L+1)J_2\right)J_1m^2T^2
\right. \\
&&\hspace{1cm}
\left.
+\left(3(3L+4)J_1^2+3J_1^2J_2+6K_2+4K_3\right)
T^4
\right]\;.
\label{f3loop}
\eqa

\subsection{Pressure to three loops}

The pressure ${\cal P}$ is given by $-{\cal F}$.
The contributions to the free energy of zeroth, first, and second order in 
$g^2$ are given in~(\ref{f1loop}),~(\ref{f2loop}), 
and~(\ref{f3loop}), respectively.
Adding them and setting ${\cal E}_0=0$ and $m_1^2=m^2$, we get the
approximations to the pressure in screened perturbation theory.
The one-loop 
approximation is obtained by  setting ${\cal E}_0=0$ in~(\ref{f1loop}):
\bqa
\label{1lop}
(4\pi)^2{\cal P}_0={1\over8}\left[
4J_0T^4+\left(2L+3\right)m^4
\right]\;.
\eqa
The two-loop approximation is obtained by 
adding~(\ref{f2loop}) with $m_1^2=m^2$:
\bqa
\label{ambi}
(4\pi)^2{\cal P}_{0+1}={1\over8}\left[
4J_0T^4+4J_1m^2T^2-\left(2L+1\right)m^4
\right]
-{1\over8}\alpha\left[
J_1T^2-(L+1)m^2
\right]^2
\;.
\eqa
The three-loop approximation is obtained by adding~(\ref{f3loop})
with $m_1^2=m^2$:
\bqa\nonumber
(4\pi)^2{\cal P}_{0+1+2}&=&{1\over8}\left[
4J_0T^4+4J_1m^2T^2+2J_2m^4-m^4
\right]
\nonumber\\ \nonumber
&&
-{1\over8}\alpha\left[J_1T^2-(L+1)m^2\right]\left[J_1T^2
+2J_2m^2+(L-1)m^2\right] 
\nonumber
\\ \nonumber
&&
+{1\over48}\alpha^2
\left[
3J_2\left(J_1T^2-(L+1)m^2\right)^2
+\left(3(3L+4)J_1^2+6K_2+4K_3\right)T^4\right.
\nonumber
\\ \nonumber
&&
\hspace{2cm} 
-\left(12L^2+28L-12-\pi^2-4C_1\right)J_1m^2T^2
\nonumber
\\  
&&
\hspace{2cm} 
+\left( 5L^3+17L^2+\mbox{${41\over2}$}L-23-\mbox{${23\over12}$}\pi^2
	-\psi^{\prime\prime}(1)+ C_0 \right)m^4
\label{freefin}
\Big]\;,
\eqa
where $L=\log(\mu^2/m^2)$, $C_0=39.429$, $C_1=-9.8424$,
the $J_n$'s are the functions of $m/T$ given in~(\ref{jn2}),
and $K_2$ and $K_3$ are functions of $m/T$ given in Ref.~\cite{massive}.
Note that the dependence on $L$ has canceled from the term
proportional to $\alpha^0$ in~(\ref{freefin}).

\subsection{Entropy to Three Loops}

The perturbative expansion for the entropy density ${\cal S}$ is defined
in~(\ref{s}). The one-, two-, and three-loop approximations 
to ${\cal S}$ are obtained by taking the partial derivatives with respect to 
$T$, with $\alpha$, $m$, and $\mu$ fixed, of the expressions for the
pressure in~(\ref{1lop}),~(\ref{ambi}), and~(\ref{freefin}).
The partial derivatives of the functions $J_n(\beta m$) 
can be evaluated using the
recursion relation~(\ref{rec}). The partial derivatives of $K_n(\beta m)$
can be evaluated numerically.

The one-loop approximation is obtained by differentiating~(\ref{1lop}):
\bqa
\label{1s}
{(4\pi)^2}T{\cal S}_0=2J_0T^4+J_1m^2T^2\;.
\eqa
The two-loop approximation is obtained by differentiating~(\ref{ambi}):
\bqa
\label{2s}
{(4\pi)^2}T{\cal S}_{0+1}=\left[2J_0T^4+2J_1m^2T^2
+J_2m^4\right]-{1\over2}\alpha\left[J_1T^2-(L+1)m^2
\right]
\left[J_1T^2+J_2m^2\right]\;.
\eqa
The three-loop approximation is obtained by differentiating~(\ref{freefin}):
\bqa
(4\pi)^2T{\cal S}_{0+1+2}&=&
{1\over2}\left[4J_0T^4+4J_1m^2T^2+2J_2m^4+J_3m^6T^{-2}\right]
\nonumber
\\ 
&& -{1\over2} \alpha 
\left[ \left( J_1T^2+J_2m^2 \right)^2 
	- \left( J_1T^2+J_2m^2 \right) m^2
\right.
\nonumber
\\
&& \left. \hspace{2cm} 
	+ J_3 \left(J_1T^2-(L+1)m^2\right) m^4 T^{-2} \right] 
\nonumber
\\ \nonumber
&&+{1\over24}\alpha^2
\left[3J_3\left(J_1T^2-(L+1)m^2\right)^2m^2T^{-2}
\right. \\ \nonumber
&&\left.
\hspace{2cm} 
+ 6 J_2 \left(J_1T^2-(L+1)m^2\right) \left( J_1T^2+J_2m^2 \right)
\right.
\\ \nonumber
&&
\hspace{2cm} 
+ \left( 6(3L+4)J_1^2 + 12 K_2 + 8K_3 \right) T^4
	- \left( 3 K_2^{\prime} + 2 K_3^{\prime} \right) m T^3
\\ \nonumber
&& 
\left. \hspace{2cm} 
	+6(3L+4)J_1J_2m^2T^{2}
\right.
\\
&&
\label{3sss}
\hspace{2cm} \left.
-\left(12L^2+28L-12-\pi^2-4C_1\right) \left(J_1T^2+J_2m^2\right) m^2\right]\;.
\eqa
The primes on $K_2$ and $K_3$ denote differentiation with respect to 
$\beta m$.

\section{Screening Mass to Two Loops}
\label{masscal}

In this section, we calculate the screening mass to two loops. 
The diagrams for the self energy that are included at this order 
are those shown in Fig.~\ref{2ldiagrams} 
together with diagrams involving counterterms. The screening
mass $m_s$ is the solution to the equation~(\ref{scrdef}).
This equation can be solved order-by-order in powers of $\alpha$ and $m_1^2$.
The solution at zeroth order in $g^2$ is simply $m_s^2=m^2$.

\begin{figure}[htb]
\epsfysize=2.5cm
\centerline{\epsffile{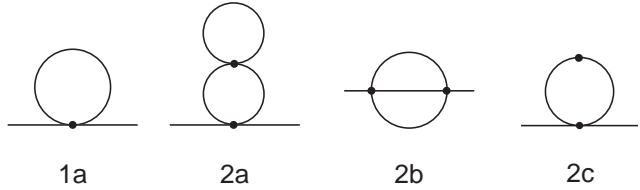}}
\vspace{3mm}
\caption[a]{Diagrams for the one-loop (1a and 1b) and two-loop (2a, 
2b and 2c) self-energy.}
\label{selfdiagrams}
\end{figure}

\subsection{One-loop self-energy}

The self-energy at first order in $g^2$ is 
\bqa
\label{g1}
\Pi_1=\Pi_{\rm 1a}-m_1^2+\Delta_1m^2\;,
\eqa
where $\Delta_1m^2$ is the mass counterterm of order $\alpha$ given 
in~(\ref{dmm}). The expression for the diagram 1a in Fig.~\ref{selfdiagrams} is
\bqa\nonumber
\Pi_{\rm 1a}
&=&{1\over2}g^2\sumint_P{1\over P^2+m^2}\;.
\eqa
The result for the diagram is
\bqa
\Pi_{\rm 1a}&=&
{1\over2}\alpha\left({\mu\over m}\right)^{2\epsilon}
\Bigg\{-\left[{1\over\epsilon}+1+{12+\pi^2\over12}\epsilon
\right]m^2+J_1T^2\Bigg\}\;.
\label{1lres}
\eqa
We have kept all terms that contribute to order $\epsilon$, because they are 
needed for counterterm diagrams in the two-loop self-energy.
The pole in $\epsilon$ in~(\ref{1lres}) 
is canceled by the counterterm $\Delta_1m^2$.
The final result for the one-loop self-energy is
\bqa
\Pi_{1}={1\over2}\alpha\left[J_1T^2-(L+1)m^2\right]-m^2_1\;.
\label{pi1}
\eqa

\subsection{Two-loop self-energy}

The contribution to the self-energy of second order in $g^2$ is
\bqa
\label{2scree}
\Pi_{2}(P)=\Pi_{\rm 2a}+\Pi_{\rm 2b}(P)+\Pi_{\rm 2c}
+{\partial\Pi_{\rm 1a}\over\partial m^2}\Delta_1m^2
+{\Pi_{\rm 1a}\over g^2}\Delta_1g^2+\Delta_2m^2
-\Delta_1m_1^2\;.
\eqa
The expressions for the diagrams 2a and 2b in Fig.~\ref{selfdiagrams} are
\bqa
\label{2la}
\Pi_{\rm 2a}&=&-{1\over4}g^4\sumint_{Q}{1\over Q^2+m^2}
\sumint_R{1\over(R^2+m^2)^2}\;,\\
\label{2lb}
\Pi_{\rm 2b}(P)&=&-{1\over6}g^4\sumint_{QR}{1\over Q^2+m^2}{1\over R^2+m^2 }{1\over (P+Q+R)^2+m^2}\;,\\
\label{2lc}
\Pi_{\rm 2c}&=&{1\over2}g^2m_1^2\sumint_{Q}{1\over(Q^2+m^2)^2}\;.
\eqa
The diagrams $\Pi_{\rm 2a}$ and $\Pi_{\rm 2c}$ are independent of the momentum $P$. 
The results for these diagrams in the limit $\epsilon\rightarrow 0$ are
\bqa
\label{2ade}
\Pi_{\rm 2a}&=&
{1\over4}\alpha^2
\left({\mu\over m}\right)^{4\epsilon}
\left[\left({1\over\epsilon^2}+{1\over\epsilon}+{6+\pi^2\over6}\right)m^2
+\left({1\over\epsilon}+1\right)J_2m^2
-\left({1\over\epsilon}+J_2\right)J_1T^2
\right] \;,\\
\label{2cde}
\Pi_{\rm 2c}&=&{1\over2}\alpha m^2_1\left({\mu\over m}\right)^{2\epsilon}
\left[{1\over\epsilon}+J_2\right]\;.
\eqa
The diagram $\Pi_{\rm 2b}$ depends on the external momentum $P$.
The equation~(\ref{scrdef}) for the screening mass involves the self-energy
at $p_0=0$. To calculate the screening mass to second order in $g^2$, 
we need the analytic  continuation 
of $\Pi(0,{\bf p})$ to ${\bf p}^2=-m^2$.
This is calculated in the appendix. The result is
\bqa
\label{2bde}
\Pi_{\rm 2b}(0,{\bf p})\Bigg|_{{\bf p}^2=-m^2}=
{1\over6}\alpha^2\left({\mu\over m}\right)^{4\epsilon}
\Bigg\{
\left[{3\over2\epsilon^2}+{17\over4\epsilon}-C_1\right]m^2
-3\left[
{1\over\epsilon}J_1+\tilde{K}_1+\tilde{K}_2
\right]T^2
\Bigg\}\;.
\eqa
The poles in~(\ref{2ade})-(\ref{2bde}) are canceled by the counterterms
in~(\ref{2scree}). 
The final result for the two-loop self-energy at $p_0=0$ and ${\bf p}^2=-m^2$
is
\bqa\nonumber
\Pi_{2}(0,{\bf p})\Bigg|_{{\bf p}^2=-m^2}&=&
{1\over2}\alpha(L+J_2)m_1^2
+{1\over24}\alpha^2\Bigg\{\left[
12L^2+28L-12-\pi^2-4C_1+
6\left(L+1\right)J_2\right]m^2
\\ 
&&
\label{scree2}
\hspace{4.16cm}
-6\left[\left(3L+J_2\right)J_1+2\tilde{K}_1+2\tilde{K}_2\right]T^2
\Bigg\}\;.
\eqa

\subsection{Screening mass}

Since the dependence of the self-energy on the momentum enters only at
order $g^4$ and since the leading-order solution to the screening mass is 
$m_s=m$, the solution to the equation~(\ref{scrdef}) 
to order $g^4$ is simply
\bqa
\label{screexpl}
m_s^2=m^2+\Pi(0,{\bf p}^2)\Bigg|_{{\bf p}^2=-m^2}\;.
\eqa
We proceed to calculate the expression to order $g^2$ and to order $g^4$.

The solution to order $g^2$ is obtained by inserting the one-loop 
self-energy~(\ref{pi1}) into~(\ref{screexpl}). Setting $m_1^2=m^2$, the
result is
\bqa
\label{ss1}
m_s^2={1\over2}\alpha\left[J_1T^2-(L+1)m^2\right]\;.
\eqa
If we choose $m=m_s= m_*$, this is identical to the one-loop 
gap equation~(\ref{pet}).

The solution to order $g^4$ is obtained by inserting the sum 
of~(\ref{pi1}) and~(\ref{scree2}) into~(\ref{screexpl}).
Setting $m_1^2=m^2$, the result is
\bqa\nonumber
m_s^2&=&{1\over2}\alpha\left[J_1T^2+(J_2-1)m^2\right]
-{1\over24}\alpha^2\Bigg[
6J_2\left(J_1T^2-(L+1)m^2\right)
\\
&&
+6\left(3LJ_1+2\tilde{K}_1+2\tilde{K}_2\right)T^2
-\left(
12L^2+28L-12-\pi^2-4C_1
\right)m^2
\Bigg]
\;.
\label{twos}
\eqa
Note that the dependence on $L$ has canceled in the order-$\alpha$ terms.

\section{Gap Equations}
\label{gapeqs}

In this section, we solve the gap equations that determine the arbitrary mass
parameter in screened perturbation theory.
We consider the one-loop gap equation and three generalizations
to a two-loop gap equation.

\subsection{One-loop Gap Equation}

The one-loop gap equation is given in~(\ref{pet}). 
It is convenient to introduce the gap function defined by 
\bqa
{\rm G}=m^2-{1\over2}\alpha\left[J_1T^2-(L+1)m^2\right]\;.
\eqa
The one-loop gap equation is then ${\rm G}=0$.
For simplicity of notation, we will often suppress the subscripts $*$ on $m$
and $\mu$.

Before solving the one-loop gap 
equation, we need to choose a value for $\mu$. It is natural to take
$\mu$ to be proportional to one of the two energy scales in the equation,
$T$ and $m$. We will consider two possibilities, $\mu=a(2\pi T)$ and
$\mu=am$, and allow the coefficient $a$ to vary from ${1\over2}$ to $2$.
Given either of these choices for $\mu$, the gap equation can be solved
for $m$ as a function of $\alpha(\mu)$.
The renormalization group equation~(\ref{rg2}) can then be used to express
$\alpha(\mu)$ as a function of $\alpha(2\pi T)$.

In the weak-coupling limit $g\rightarrow 0$, the solution to
the gap equation G = 0 approaches
\bqa
\label{petlim}
m_*^2 \longrightarrow {2\pi^2\over3}\alpha(\mu)T^2
\left[ 1 - \sqrt{6} \alpha^{1/2} 
- \left(\log{\mu\over4\pi T} + \gamma - 3 \right) \alpha 
+ O(\alpha^{3/2}) \right] \, .
\eqa
In the strong-coupling limit $g\rightarrow\infty$, the gap equation
reduces to
\bqa
\label{redgap}
2\log{\mu\over m}+1=8\int_0^{\infty}
dx\;{x^2\over\sqrt{1+x^2}}{1\over e^{\beta m\sqrt{1+x^2}}-1}\;.
\eqa
This has a solution only if $\mu>e^{-1/2}m$. 

\begin{figure}[htb]
\epsfysize=8cm
\centerline{\epsffile{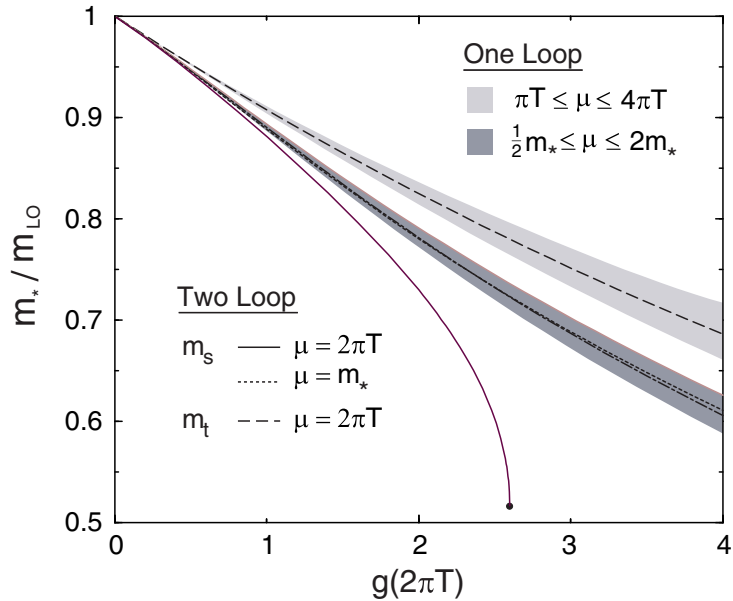}}
\vspace{3mm}
\caption[a]{Solutions $m_*(T)$ to the one-loop gap equation (shaded bands)
	and the two-loop gap equations (lines) as functions of $g(2\pi T)$.}
\label{msgap1}
\end{figure}

In Fig.~\ref{msgap1}, the solutions $m_*(T)$ to the one-loop gap 
equation as a function of $g(2\pi T)$ are shown as bands obtained by 
varying $\mu$ by a factor of two around the central values $\mu=2\pi T$
and $\mu=m_*$, respectively.  The solutions are
normalized to the leading-order screening mass 
$m_{LO}=g(2\pi T)T/\sqrt{24}$.

\subsection{Screening Gap Equation}

The screening gap
equation is obtained by identifying $m$ with $m_s$.
The one-loop expression for the screening mass is given in~(\ref{ss1}).
Thus the one-loop screening gap equation is simply ${\rm G}=0$.
The two-loop expression for the screening mass is given in~(\ref{twos}).
The two-loop screening gap equation can be written as 
\bqa\nonumber
\left[1-{1\over2}\alpha\left(
J_2+L
\right)\right]{\rm G}
+{1\over24}\alpha^2\bigg[
12\left(LJ_1+\tilde{K}_1+\tilde{K}_2
\right)T^2&& \\ 
-\left(
6L^2+22L-12 -\pi^2 -4C_1
\right)m^2\bigg]
&=&0\;.
\eqa
From this expression, it is easy to see that the solution $m$
to the gap equation
differs from the solution~(\ref{petlim}) to the one-loop gap equation
by terms of order $\alpha^2T^2$. The weak-coupling expansion of the
solution $m^2$ must of course agree through order $\alpha^2T^2$
with the weak-coupling expansion
of $m_s^2$ given in~(\ref{mpert}).

The solutions to the screening gap equation for $\mu=2\pi T$ and $\mu=m_*$
are shown in Fig.~\ref{msgap1}. In the case
$\mu=2\pi T$, the screening gap equation 
cannot be continued beyond $g(2 \pi T)=2.60$.
For $\mu=\pi T$, it terminates at $g(2 \pi T)=2.31$, while 
for $\mu=4\pi T$, it terminates at $g(2 \pi T)=3.04$.
If we choose $\mu=am_*$, the solution can be 
continued to much larger values of $g$.
For $\mu=m_*$, it lies very close to the solution 
to the one-loop gap equation with $\mu = m_*$.

\subsection{Tadpole Gap Equation}

The tadpole mass $m_t$ is defined in~(\ref{diffdef}). 
The one-loop expression
is given by differentiating~(\ref{f1loop}). The result is
identical to the one-loop expression~(\ref{ss1}) for the screening mass.
To obtain the two-loop expression for the tadpole mass, 
we add the one- and two-loop free energies~(\ref{f1loop}) 
and~(\ref{f2loop}), differentiate  with respect to $m^2$, 
and then set $m_1^2=m^2$.
The result is
\bqa
\label{mt*}
m_t^2={1\over2}\alpha\left[J_1T^2+(J_2-1)m^2
\right]
-{1\over4}\alpha^2\left(J_2+L\right)\left[J_1T^2-(L+1)m^2\right]\;.
\eqa
The order-$\alpha$ term is identical to that of the screening 
mass~(\ref{twos}), but the order-$\alpha^2$ term is much simpler.

The one-loop tadpole equation is simply ${\rm G}=0$.
The two-loop tadpole gap equation is obtained by setting $m_t=m$ 
in~(\ref{mt*}).  It can be written in the form
\bqa
\left[1-{1\over2}\alpha\left(J_2+L\right)\right]{\rm G}=0\;.
\eqa
Thus the two-loop tadpole gap equation is identical to the one-loop
gap equation: ${\rm G}=0$.
The solutions for $\mu = 2 \pi T$ and $\mu = m_*$ are at the 
centers of the shaded bands in Fig.~\ref{msgap1}.

\subsection{Variational Gap Equation}

The variational mass $m_v$ is the solution to~(\ref{vmass}).
The one-loop variational gap equation is obtained by differentiating
the two-loop expression (\ref{ambi}) for the pressure
with respect to $m^2$ and setting it equal to zero. 
This gives $\left(L+J_2\right)m^2 {\rm G}=0$, 
which reduces to the one-loop gap equation: ${\rm G}=0$.

The two-loop variational gap equation is
obtained by differentiating the three-loop expression (\ref{freefin})
for the pressure.
It can be expressed in the form
\bqa \nonumber
0&=&
{1\over4}\alpha(J_2+L)^2{\rm G}-{1\over4}\left(
J_3+{1\over(\beta m)^2}\right)
{\rm G}^2/T^2 
\\ 
\label{v*}
&& \hspace{2mm} +{1\over48}\alpha^2
\left[
-6{J_1^2\over(\beta m)^2}-12(L+2)J_1J_2
+{3K_2^{\prime}+2K_3^{\prime}\over\beta m}+ \;...\;
\right]T^2,
\eqa
where $K_2^{\prime}$ and $K_3^{\prime}$ are the derivatives of $K_2$ and $K_3$
with respect to $\beta m$.
In the coefficient of $\alpha^2T^2$, we have written explicitly only the
terms that are singular as $\beta m\longrightarrow 0$.
The $1/(\beta m)^2$ singularities cancel between the $J_1^2$ and 
$K_2^{\prime}$ term. If we keep the most singular terms in the coefficients
of each of the three terms in~(\ref{v*}), the equation reduces to
\bqa
\label{v**}
0 &=&{\pi^2\alpha\over(\beta m)^2}{\rm G}
-{\pi\over4(\beta m)^3}{\rm G}^2/T^2
\nonumber
\\
&& \hspace{2mm} - \left[ 32 \pi^3 (L+2) 
	- (3 k_2' + 2 k_3') \log(\beta m)
	- 3(k_2 + k_2') - 2(k_3 + k_3') \right]
	{\alpha^2 T^2\over 48 \beta m}=0\;,
\eqa
where $k_2'\log(\beta m) + k_2$ and $k_3' \log(\beta m) + k_3$
are the coefficients of $\beta m$ in the small-$\beta m$
expansions of $K_2$ and $K_3$, which are given in (\ref{K2}) and (\ref{K3}).

The solution to the quadratic equation~(\ref{v**}) for
${\rm G}$ is proportional to $\alpha \beta m T^2$. The solution $m^2$
to the gap equation therefore differs from the solution~(\ref{petlim}) 
to the one-loop gap equation by terms of order $\alpha^{3/2}T^2$.
This is a little disturbing, but even more disturbing is the fact 
that~(\ref{v**}) has no real-valued solutions for G unless 
$L < 2.0984 \log(\beta m) + 4.1541$.
If we assume that $m \to gT/\sqrt{24}$ as $g \to 0$,
then this condition is violated for sufficiently small $g$
whether we set $\mu=a(2 \pi T)$ or $\mu=am$.
Since there are no solutions 
in the neighborhood of $g=0$, we will not consider 
the two-loop variational gap equation any further.

\section{SPT-improved Observables}
\label{observables}

In this section, we use the solutions to the gap equation in Sec.~\ref{gapeqs} to
obtain successive approximations to the pressure, screening mass, and
entropy
in screened perturbation 
theory.

\subsection{Pressure}
\begin{figure}[htb]
\epsfysize=8cm
\centerline{\epsffile{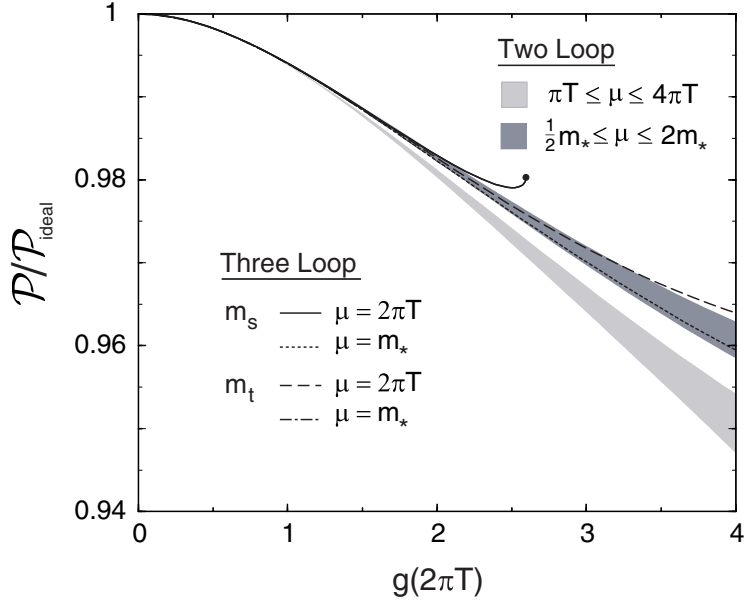}}
\vspace{3mm}
\caption[a]{Two-loop (shaded bands) and three-loop (lines)
SPT-improved pressure as a function of $g(2\pi T)$.}
\label{freeplot1}
\end{figure}

The two-loop SPT-improved
approximation to the pressure  is obtained by inserting the
solution to the one-loop gap equation~(\ref{pet}) into the two-loop 
pressure~(\ref{ambi}).
We can simplify the expression by using~(\ref{pet}) to eliminate the explicit
appearance of logarithms of $\mu$.
Remarkably, this eliminates all the terms of order $\alpha$ and the expression
reduces simply to
\bqa
(4\pi)^2{\cal P}_{0+1} &=& 
{1\over8}\left[4J_0T^4+2J_1m^2T^2+m^4
\right]\;.
\label{P-01}
\eqa
The $J_0$ term in (\ref{P-01}) is the pressure 
of an ideal gas of particles of mass $m$.
Inserting the solution to the one-loop gap equation shown in 
Fig.~\ref{msgap1}, we
obtain the bands shown in Fig.~\ref{freeplot1}. 
The lower and upper bands correspond to
varying $\mu$ by a factor of $2$ around the central values $\mu=2\pi T$
and $\mu=m_*$, respectively.

The three-loop SPT-improved approximation to the pressure is obtained by
inserting the solution to a two-loop gap equation into the three-loop 
pressure~(\ref{freefin}). 
In Fig.~\ref{freeplot1}, we show the three-loop SPT-improved pressure
as a function of $g(2\pi T)$ for different two-loop gap equations. 
The solid line is the result using the two-loop screening gap equation with 
$\mu=2\pi T$.  It cannot be extended past $g(2 \pi T)=2.60$. 
The dashed line is the result using the two-loop tadpole 
(or one-loop) gap equation with $\mu=2\pi T$.
The dotted line is the result using either the two-loop screening gap equation 
with $\mu_*=m_*$ or the two-loop tadpole gap equation with $\mu_*=m_*$.
The two are indistinguishable on the scale of the figure. 
The variations among the three-loop SPT-improved approximations 
for the pressure are much smaller than one might have expected from the 
variations among the screening masses.
For example, at $g(2 \pi T) = 2$, the solutions to the two-loop gap 
equations shown in Fig.~\ref{msgap1} vary by about 12\%, 
while the three-loop approximations to the pressure shown 
in Fig.~\ref{freeplot1} vary only by about 0.07\%.

\begin{figure}
\epsfysize=7cm
\centerline{\epsffile{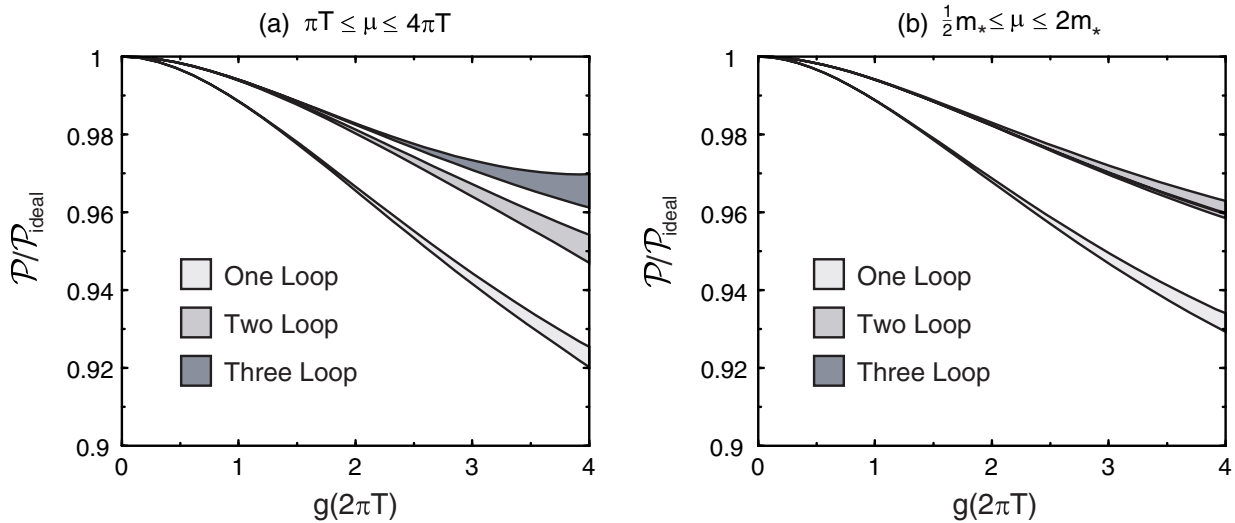}}
\vspace{3mm}
\caption[a]{One-, two-, and three-loop SPT-improved pressure
	as a function of $g(2\pi T)$ for
	(a) $\pi T < \mu < 4 \pi T$ and 
	(b) ${1 \over 2} m_* < \mu < 2 m_*$.
}
\label{freeplot2}
\end{figure}

Since the solution to the screening gap equation at $\mu = a(2 \pi T)$ 
cannot be continued beyond a critical value of $g$ 
and the solution for $\mu = a m_*$ is close to the solution to the 
tadpole gap equation for $\mu = a m_*$, we will consider only the 
tadpole gap equation from now on.
In Fig.~\ref{freeplot2}, we show the one-, two-, and three-loop SPT-improved
approximations to the pressure using the tadpole gap 
equation. 
The bands are obtained by varying $\mu$ by a factor of two around
the central values $\mu=2\pi T$ and $\mu=m_*$.
The one-loop bands in Fig.~\ref{freeplot2} lie below the other bands;
however, the two- and three-loop bands all lie within the $g^5$ band of the
weak-coupling expansion in Fig.~\ref{fpert}.
The one-, two-, and three-loop approximations to the pressure are
perturbatively correct up to order $g^1$, $g^3$, and $g^5$, respectively; 
however, we see a dramatic improvement in the apparent convergence
compared to the weak-coupling expansion.   

The choice $\mu = a m_*$ appears to give better convergence than 
$\mu = a(2 \pi T)$, with the three-loop band falling within 
the two-loop band.  The bands for $\mu = a m_*$ are narrower than 
those for $\mu = a(2 \pi T)$ partly because $\mu = a(2 \pi T)$
is larger and therefore closer to the Landau pole of the 
running coupling constant. 
If $g(2 \pi T)=2$, the Landau pole associated with the five-loop 
beta function is far away at $\mu = 2.11\times10^5 (2 \pi T)$.
If $g(2 \pi T)=4$, the Landau pole is rather nearby at $\mu = 5.49 (2 \pi T)$.
The coupling constant $g(m_*)$ is smaller than $g(2\pi T)$, having
the values 1.76 and 3.07 if $g(2\pi T)=2$ and $4$, respectively.  Choosing
$\mu = a m_*$ instead of $\mu = a (2\pi T)$ will therefore make the error
due to the $m^4$ terms in the pressure smaller by factors of about 0.60 and
0.35 respectively.  The band $m_*/2 < \mu < 2 m_*$ may therefore give an
underestimate of the error of SPT.

\subsection{Screening Mass}

\begin{figure}
\epsfysize=7cm
\centerline{\epsffile{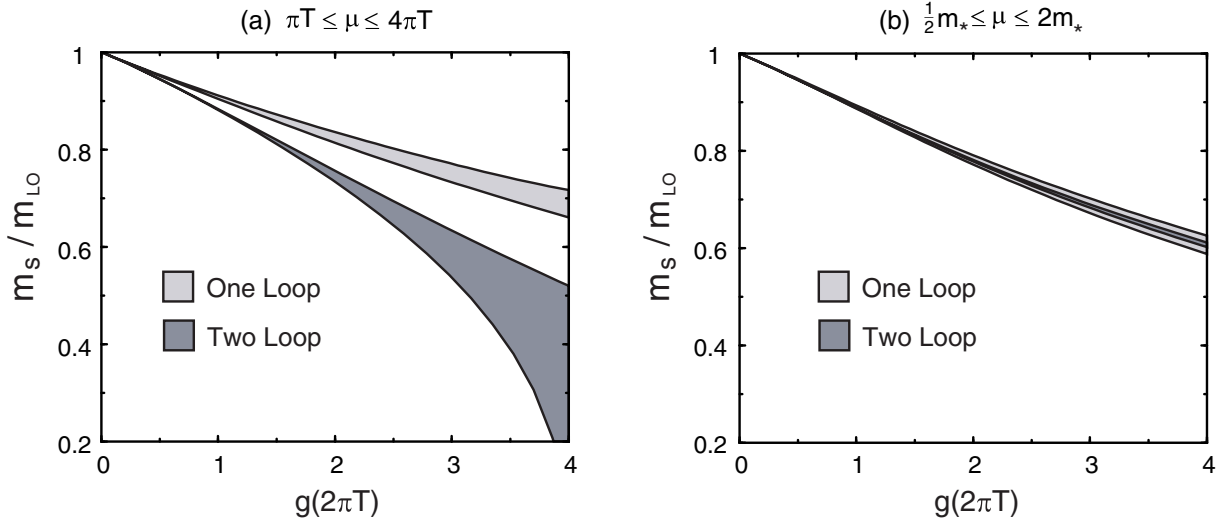}}
\vspace{3mm}
\caption[a]{One-loop and two-loop SPT-improved screening mass
	as a function of $g(2\pi T)$ for
	(a) $\pi T < \mu < 4 \pi T$ and 
	(b) ${1 \over 2} m_* < \mu < 2 m_*$.
}
\label{freeplot3}
\end{figure}

The one-loop SPT-improved approximation to the screening mass
$m_s$ is simply the solution $m_*(T)$ to the tadpole gap equation.
A two-loop SPT-improved approximation can be obtained by inserting the
solution to the gap equation for $m$ into~(\ref{twos}).
In Fig.~\ref{freeplot3}, we show the one-loop and two-loop SPT-improved
approximations to the screening mass as functions of $g(2\pi T)$.
The bands are obtained by varying $\mu$ by a factor of two around
the central values $\mu=2\pi T$ and $\mu=m_*$.

The choice $\mu = a m_*$ appears again to give better convergence than 
$\mu = a(2 \pi T)$, with the two-loop band falling within 
the one-loop band.  With $\mu = a m_*$,
there is a dramatic improvement in apparent convergence over
the weak-coupling approximations, 
which are plotted on the same scale in Fig.~\ref{mspert}.
However, there is not much
improvement in the apparent convergence with $\mu = a(2 \pi T)$.
The conservative conclusion is that screened perturbation theory 
is not as effective in improving the prediction for the screening mass 
as it is for the pressure.

\subsection{Entropy}

\begin{figure}[htb]
\epsfysize=7cm
\centerline{\epsffile{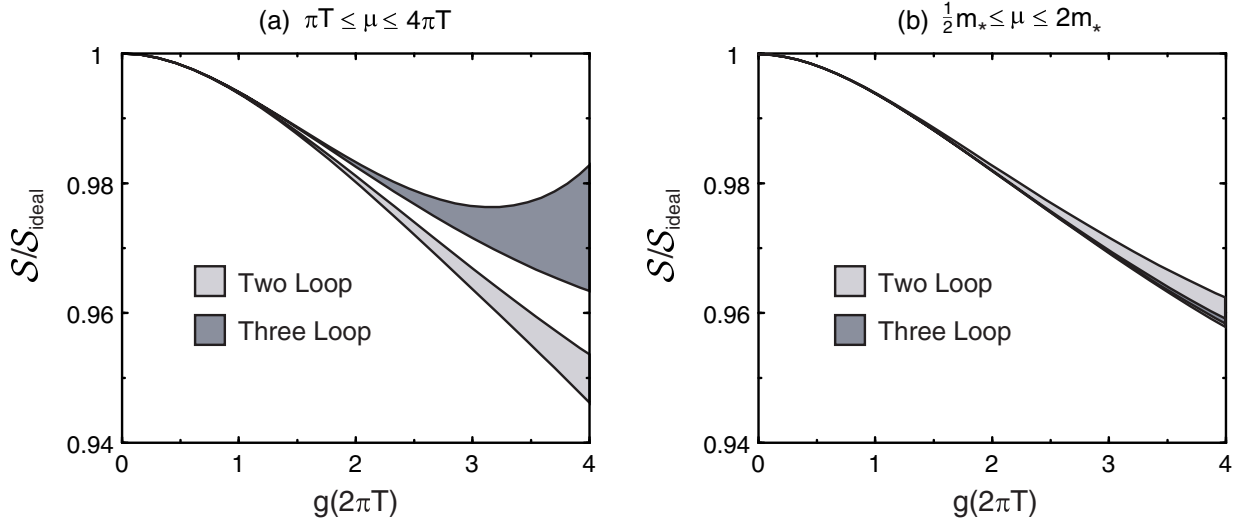}}
\vspace{3mm}
\caption[a]{Two-loop and three-loop SPT-improved entropy
	as a function of $g(2\pi T)$ for
	(a) $\pi T < \mu < 4 \pi T$,
	(b) ${1 \over 2} m_* < \mu < 2 m_*$.}
\label{2s3s}
\end{figure}

The one-, two- and three-loop SPT-improved entropies are obtained 
by replacing $m$ in the expressions~(\ref{1s})-(\ref{3sss}) 
for ${\cal S}_0$, ${\cal S}_{0+1}$, and ${\cal S}_{0+1+2}$
with the solution $m_*$ to the one-loop gap equation ${\rm G} = 0$.
Using the gap equation to eliminate the logarithm $L$, the expression
for the two-loop entropy reduces to
\bqa
{(4\pi)^2}T{\cal S}_{0+1}=2J_0T^4+J_1m^2T^2
\;.
\eqa
This is identical to the one-loop expression~(\ref{1s}), which is 
the entropy of an ideal gas of particles with mass $m$.
In Fig.~\ref{2s3s}, we show
the two- and three-loop SPT-improved  approximations to the entropy
as functions of $g(2\pi T)$. The entropy density is normalized to that
of an ideal gas: ${\cal S}_{\rm ideal}=(2\pi^2/45)T^3$.
The bands in Fig.~\ref{2s3s} correspond to varying $\mu$ by a factor of
two around the central values $2\pi T$ and $m_*$.
Once again, the choice $\mu = a m_*$ seems to give better convergence
with the three-loop band lying very close to the two-loop band.

\begin{figure}[htb]
\epsfysize=7cm
\centerline{\epsffile{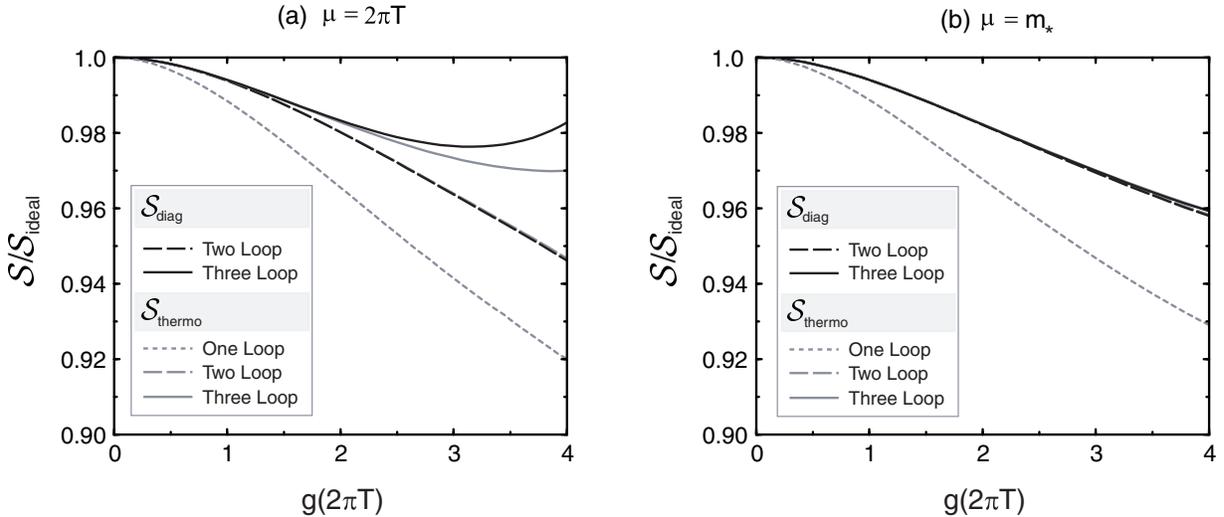}}
\vspace{3mm}
\caption[a]{SPT-improved entropy
	as a function of $g(2\pi T)$ 
	compared to the thermodynamic entropy for obtained from
	the SPT-improved pressure for
	(a) $\mu = 2 \pi T$ and 
	(b) $\mu = m_*$.}
\label{S-thermo}
\end{figure}

The entropies shown in Fig.~\ref{2s3s} are successive approximations to
the diagrammatic entropy defined by Eq.~\ref{e}.  However, the entropy
can also be defined by 
the thermodynamic relation~(\ref{s}). 
Thus successive 
approximations to ${\cal S}$ can be obtained by
differentiating the pressures shown in Fig.~\ref{freeplot2} with
respect to $T$. 
In that figure we show the ratio of the pressure to that 
of an ideal gas as a function of $g(2\pi T)$.  
Defining the function $f(g)$ by
\bqa
{\cal P}(T) = {\cal P}_{\rm ideal}(T) f(g(2\pi T)) \; ,
\eqa
the thermodynamic entropy is then given by
\bqa
{\cal S}_{\rm thermo}(T) \;=\; {\cal S}_{\rm ideal}(T) \left[ f(g) + 
	{2 \pi^2 \over g} f'(g) \beta(\alpha) \right] 
\;,
\eqa
where $g = g(2 \pi T)$, $\alpha = g^2/16\pi^2$, and $\beta(\alpha)$
is the beta function given by the right side of (\ref{rg2}).
In Fig.~\ref{S-thermo}, the black curves are the two- and three-loop 
diagrammatic entropies for $\mu = 2 \pi T$ and $\mu = m_*$.  
The gray curves are the the thermodynamic entropies
obtained from the one-, two-, and three-loop SPT-improved pressures.
One can see clearly the approach to thermodynamic consistency
as one goes from the two-loop to the three-loop approximation.
With the choice $\mu = 2 \pi T$, the two-loop entropy is almost
perfectly thermodynamically consistent.  However, this is probably
an accident because the deviations from thermodynamic entropy are 
evident in the three-loop entropy.  With $\mu = a m_*$, deviations
from thermodynamic consistency are very small for both the two- and
three-loop entropies.  This is another indication that SPT improvement
is more effective if we take the scale $\mu$ to be much lower than
$2\pi T$.

\section{Conclusions/Discussion}
\label{conclusions}

We have studied the effectiveness of 
screened perturbation theory in reorganizing the perturbation
series for a thermal scalar field theory.
We applied it to the pressure and the entropy calculated 
to three loops and to the screening mass calculated to two loops.

We considered three alternatives for generalizing the 
one-loop gap equation to two-loop order. The most useful
turned out to be the tadpole gap equation, 
which at two loops is identical to the one-loop gap equation proposed
by Karsch, Patk\'os and Petreczky.
The solution to the two-loop variational gap equation
does not match onto the one-loop gap equation in the weak-coupling limit.
The solution to the two-loop screening gap equation
cannot be extended above $g(2 \pi T) = 2.60$ 
if we choose the scale to be $\mu = 2 \pi T$.

The predictions of SPT depend on an arbitrary scale $\mu$ that arises both 
from the renormalization of the coupling constant and from the 
renormalization of ultraviolet divergences introduced by screened 
perturbation theory.  
These two effects could be separated by introducing
two renormalization scales, $\mu_3$ and $\mu_4$.  These scales would
be associated with contributions from soft and hard modes respectively as in Ref.~\cite{EJM1}.
One way disentangle the dependence on these scales would be to evaluate
the integrals as expansions in $m/T$.  We evaluated our integrals by integrating
numerically over all momenta, which precluded any separation of the scales.
Instead, we considered two possibilities for the 
scale, $\mu = m_*$ and $\mu = 2 \pi T$, which correspond to the central values
expected for $\mu_3$ and $\mu_4$, respectively.
We allowed for variations of $\mu$
around these central values by factors of two to provide a lower bound on
the theoretical uncertainty.  The choice $\mu = m_*$ gives smaller 
bands from varying the scale, but this is largely due to the fact 
that the coupling constant $g(m_*)$ is smaller than $g(2 \pi T)$.  
Thus the size of the bands is not a good indicator of the success of the SPT
improvement.

A better indication of the success of SPT improvement is the stability 
of the predictions as you go to higher order in the loop expansion.
The choice $\mu = 2 \pi T$ gives a significant improvement in stability 
for the pressure compared to the weak-coupling expansion.  
However the SPT improvement seems to be much 
more effective using $\mu = m_*$ than $\mu = 2 \pi T$.  The three-loop band lies 
within the two-loop band for the pressure and it lies very close for the 
entropy.  The two-loop band also lies within the one-loop band for the 
screening mass.  The two-loop and three-loop approximations for the 
entropy are also very close to thermodynamic consistency if we choose 
$\mu = m_*$.  If we set $\mu = 2 \pi T$, then going from the two-loop to the 
three-loop approximations to the pressure or entropy moves the
prediction closer to that for $\mu = m_*$.  This suggests that the 
SPT-improved prediction for $\mu = m_*$ is more accurate than that for
$\mu = 2 \pi T$.  All this evidence indicates that SPT improvement is most
successful if $\mu$ is taken to be much smaller than $2 \pi T$.

To remove the additional ultraviolet divergences introduced by SPT,
we have chosen to use dimensional regularization with modified
minimal subtraction.  This choice is of course not unique.
For example, we could have also chosen to subtract the 
piece of the free energy that is independent of $T$ for fixed $m$, i.e. ${\cal F}_{\rm R} = 
{\cal F}(T,g,m,\mu)-{\cal F}(T=0,g,m,\mu)$.  This would result in a different 
reorganization of the perturbation series that would also agree with the exact
result if summed to all orders.  We take into account the theoretical
uncertainty associated with the choice of subtraction scheme by allowing
variation of the renormalization scale $\mu$.  For example, by examining
Eqs.~(\ref{1lop}) and (\ref{ambi}), we can see that the alternative described
above corresponds to $L=-3/2$ or the one-loop approximation and to some
value in the range $-{1\over2}<L<-1$ for the two-loop approximation.  Setting
$\mu = a m_*$ with ${1\over2}<a<2$ corresponds to varying $L$ in the range
$-1.4 < L < 1.4$.  Therefore, this variation of $\mu$ does take into 
account the ambiguity associated with the subtraction scheme at $T=0$
It would of course be preferable to separate the ambiguity from the SPT
subtractions from the ambiguity from renormalization of the original 
theory by allowing separate renormalization scales $\mu_3$ and $\mu_4$ as
mentioned above.

Our results demonstrate the effectiveness of screened perturbation theory
in providing stable and apparently converging predictions for the thermodynamic
functions of a massless scalar field theory.  An essential ingredient of this 
approach is using the solution to a gap equation as the prescription for 
the mass parameter $m$.  This success of screened perturbation theory adds
support to the proposal of Ref.~\cite{EJM1} to use HTL perturbation theory
to reorganize the weak-coupling expansions for the thermodynamic functions
of QCD.  In Ref.~\cite{EJM1}, the free energy was computed only to one-loop
order, so there was little alternative to using a weak-coupling expression
for the thermal gluon mass parameter.  However, our experience with SPT
indicates that the stability of the predictions is greatly improved by
using a solution to a gap equation for the mass parameter.  A gap equation
can be derived from the free energy calculated to two-loop order in HTL
perturbation theory.  Until that calculation is carried out, quantitative
comparisons of the predictions of HTL perturbation theory with the
nonperturbative results of lattice gauge theory are probably premature.

\section*{Acknowledgments}
This work was supported in part by the U.~S. Department of
Energy Division of High Energy Physics
(grants DE-FG02-91-ER40690 and DE-FG03-97-ER41014)
and by a Faculty Development Grant
from the Physics Department of the Ohio State University.

\appendix
\renewcommand{\theequation}{\thesection.\arabic{equation}}

\section{Sum-integrals}

\setcounter{equation}{0}
\label{sumintegrals}

In the imaginary-time formalism for thermal field theory, a boson has
Euclidean four-momentum $P=(p_0,{\rm\bf p})$, with $P^2=p_0^2+{\rm\bf p}^2$. The
Euclidean energy $p_0$ has discrete values: $p_0=2\pi nT$, where $n$ is
an integer. Loop diagrams involve sums over $p_0$ and integrals over
{\bf p}. 
We use dimensional regularization to regularize ultraviolet or
infrared divergences. Our choice for the measure
in the sum-integrals is
\bqa
  \sumint_P \;\equiv\;
  \left(\frac{e^\gamma\mu^2}{4\pi}\right)^\epsilon\;
  T\sum_{p_0}\:\int {d^{3-2\epsilon}p \over (2 \pi)^{3-2\epsilon}}\,,
\eqa
where $3-2\epsilon$ is the dimension of space and $\mu$ is an arbitrary
momentum scale. The factor $(e^\gamma/4\pi)^\epsilon$
is introduced so that, after minimal subtraction of the poles in $\epsilon$
due to ultraviolet divergences, $\mu$ coincides with the renormalization
scale of the $\overline{\rm MS}$ renormalization scheme.

\subsection{One-loop sum-integrals}

The one-loop sum-integrals that appear in the free energy can be separated
into a temperature-independent term and a term that depends explicitly on $T$:
\bqa
\label{s1}
  \sumint_P\log\left(P^2+m^2\right)&=&
{1\over(4\pi)^2}\left({\mu\over m}\right)^{2\epsilon}
\left[-{e^{\gamma\epsilon}\Gamma(1+\epsilon)
	\over\epsilon(1-\epsilon)(2-\epsilon)}m^4-J_0T^4
\right]\;,\\
\label{s2}
\sumint_P\frac{1}{P^2+m^2}&=&
{1\over(4\pi)^2}\left({\mu\over m}\right)^{2\epsilon}
\left[-{e^{\gamma\epsilon}\Gamma(1+\epsilon)\over\epsilon(1-\epsilon)}m^2
+J_1T^2\right]\;,
\\
\label{s3}
\sumint_P\frac{1}{(P^2+m^2)^2}&=&
{1\over(4\pi)^2}\left({\mu\over m}\right)^{2\epsilon}
\left[{e^{\gamma\epsilon}\Gamma(1+\epsilon)\over\epsilon}
+J_2\right]\;.
\eqa
The thermal terms can be expressed as integrals involving the Bose-Einstein
distribution function:
\bqa
\label{jndef}
J_n(\beta m)&=&
{4e^{\gamma\epsilon}\Gamma({1\over2})\over\Gamma({5\over2}-n-\epsilon)}
\beta^{4-2n}m^{2\epsilon}
\int_0^{\infty}dk\;{k^{4-2n-2\epsilon}\over\left(k^2+m^2\right)^{1/2}}
{1\over e^{\beta\left(k^2+m^2\right)^{1/2}}-1}\;. 
\eqa
These integrals satisfy the recursion relation
\bqa
\label{rec}
xJ_n^{\prime}(x)=2\epsilon J_n(x)-2x^2J_{n+1}(x)\;.
\eqa
The temperature-independent terms in~(\ref{s1})--(\ref{s3})
can be expanded as a Laurent series around $\epsilon=0$ by using
\bqa
e^{\gamma\epsilon}\Gamma(1+\epsilon)=
1+{\pi^2\over12}\epsilon^2+{1\over6}\psi^{\prime\prime}(1)\epsilon^3
+{\cal O}(\epsilon^4)\;.
\eqa
The functions $J_n$ have Taylor expansions around $\epsilon=0$.
They often appear multiplied by poles in $\epsilon$, but
it is counterproductive to expand $J_n$ in powers of
$\epsilon$, because the poles always cancel in physical quantities.

If we set $\epsilon =0$, the integrals $J_n$ for $n=0,1,2$  reduce to
\bqa
\label{jn2}
J_n(\beta m)={4\Gamma({1\over2})\over\Gamma({5\over2}-n)}\beta^{4-2n}
\int_0^{\infty}dk\;{k^{4-2n}\over(k^2+m^2)^{1/2}}
{1\over e^{\beta(k^2+m^2)^{1/2}}-1}\;.
\eqa 
The integral $J_3$ requires a subtraction to remove a linear infrared
divergence:
\bqa
J_3(\beta m)=
-2\beta^{-2}\int_0^{\infty}dk\;{1\over k^2}
\left(
{1\over\left(k^2+m^2\right)^{1/2}}{1\over e^{\beta(k^2+m^2)^{1/2}}-1}
-{1\over m}{1\over e^{\beta m}-1}
\right)\;.
\eqa
In the limit $\beta m\longrightarrow 0$, these
integrals reduce to
\bqa
J_0 &\longrightarrow& {16\pi^4\over45}\;,\\
J_1 &\longrightarrow& {4\pi^2\over3}
	\;-\; 4\pi \beta m
	\;-\; 2 \left( \log{\beta m\over4\pi}-{1\over2}+\gamma\right) 
		(\beta m)^2 \;,\\
J_2 &\longrightarrow& {2\pi\over\beta m}
	\;+\; 2 \left( \log{\beta m\over4\pi}+\gamma\right)\;,\\
J_3 &\longrightarrow&{\pi\over(\beta m)^3}
	\;-\; {1\over(\beta m)^2} \;+\; {1\over4\pi^2}
\zeta(3)\;.
\eqa

\subsection{Basketball sum-integral}

The only nontrivial sum-integral required to calculate the free energy
to three loops is the massive basketball sum-integral:
\bqa
{\cal I}_{\rm ball}=
\sumint_{PQR}{1\over(P^2+m^2)(Q^2+m^2)(R^2+m^2)[(P+Q+R)^2+m^2]}\;.
\eqa
This sum-integral was evaluated in Ref.~\cite{bugrij} and the result is 
\bqa\nonumber
{\cal I}_{\rm ball}&=&{1\over(4\pi)^6}\left({\mu\over m}\right)^{6\epsilon}
\Bigg\{\left[
{2\over\epsilon^3}+{23\over3\epsilon^2}+{35+\pi^2\over2\epsilon}+C_0\right]m^4
+
\left[
-{6\over\epsilon^2}-{17\over\epsilon}+4C_1
\right]J_1m^2T^2\\
&&
\hspace{2.83cm}+\left(
{6\over\epsilon}+12\right)J_1^2T^4+\left(6K_2
+4K_3\right)T^4\Bigg\}T^4\;,
\label{balldef}
\eqa
where $C_0=39.429$, $C_1=-9.8424$, and 
$K_2$  and $K_3$ are functions of $\beta m$.
They are expressed in Ref.~\cite{massive} as three-dimensional integrals
that can be evaluated numerically. Their behavior in the limit
$\beta m \to 0$ is
\bqa
K_2 &\longrightarrow&
{32\pi^4\over9} \left[ \log(\beta m) - 0.04597 \right]
\;-\; 372.65 \left[ \log(\beta m) + 1.4658 \right] \beta m \;,
\label{K2}
\\
K_3 &\longrightarrow&
453.51 
\;+\; 1600.0 \left[ \log(\beta m) + 1.3045 \right] \beta m\;.
\label{K3}
\eqa
The leading terms are given analytically in Ref.~\cite{massive}.
The terms proportional to $\beta m$ were determined numerically.

\subsection{Sunset sum-integral}

The only nontrivial sum-integral required to calculate
the self-energy to two loops is the sunset
sum-integral, which depends on the external four-momentum $P=(p_0,{\bf p})$:
\bqa
{\cal I}_{\rm sun}(P)=
\sumint_{QR}{1\over Q^2+m^2}{1\over R^2+m^2}{1\over (P+Q+R)^2+m^2}\;.
\eqa
This sum-integral can be separated into terms with zero, one, and two
thermal distributions, respectively~\cite{B-J-S}. At $p_0=0$, 
it can be written as
\bqa
\label{sep}
{\cal I}_{\rm sun}(0,{\bf p})={\cal I}_{\rm sun}^{(0)}({\bf p}^2)
+3{\cal I}_{\rm sun}^{(1)}({\bf p}^2)+3{\cal I}_{\rm sun}^{(2)}({\bf p}^2)\;,
\eqa
where 
\bqa
{\cal I}_{\rm sun}^{(0)}({\bf p}^2) &=&
\int_{QR}{1 \over Q^2+m^2} {1 \over R^2+m^2} 
	{1 \over (P+Q+R)^2+m^2}\Bigg|_{P=(0,{\bf p)}} \;,
\label{I0-def}
\\
{\cal I}_{\rm sun}^{(1)}({\bf p}^2) &=&
\mbox{Re}\int_q n \delta(q)
\int_{R}{1 \over R^2+m^2}
	{1 \over (P+Q+R)^2+m^2}\Bigg|_{(P+Q)^2 = -[E_q^2-({\bf p}+{\bf q})^2+i\epsilon]} \;,
\label{I1-def}
\\
{\cal I}_{\rm sun}^{(2)}({\bf p}^2) &=&
{\rm Re}\int_q n \delta(q) \int_r n \delta(r)
{(-1)\over (p+q+r)^2-m^2+i\epsilon}\Bigg|_{p=(0,{\bf p)}} \;.
\label{I2-def}
\eqa
The integral $\int_q$ denotes the dimensionally regularized integral over the
Minkowski momentum $(q_0,{\bf q})$, and $n\delta(q)=n(q_0)2\pi\delta(q^2-m^2)$.

\subsubsection{Zero thermal factors}

To calculate the screening mass to two loops, we need 
the analytic continuation of the integrals~(\ref{I0-def})-(\ref{I2-def}) 
to the point ${\bf p}^2=-m^2$. The
integral~(\ref{I0-def}) was evaluated in Ref.~\cite{massive}:
\bqa
\label{sc0}
{\cal I}_{\rm sun}^{(0)}(-m^2)&=&
{1\over(4\pi)^4}\left({\mu\over m}\right)^{4\epsilon}
\left[-{3\over2\epsilon^2}-{17\over4\epsilon}+C_1\right]m^2\;,
\eqa
where $C_1=-9.8424$.

\subsubsection{One thermal factor}

The integral~(\ref{I1-def}) can expressed as 
\bqa\nonumber
{\cal I}_{\rm sun}^{(1)}({\bf p}^2)&=&{1\over(4\pi)^4}
\left({\mu\over m}\right)^{4\epsilon}
\left[
{1\over\epsilon}J_1T^2
\right.
\\ 
&&\left.
\label{11f1}
-8\int_0^{\infty}dq\;{q^2n(E_q)\over E_q}
\int_0^1dx\left\langle
\log{\left|m^2-x(1-x)(E_q^2-k^2)\right|\over m^2}\right\rangle
\right]\;,
\eqa
where $k=|{\bf p}+{\bf q}|$ and $\langle...\rangle$ denotes the 
angular average.
After averaging over angles, (\ref{11f1}) can be analytically continued
to ${\bf p}^2=-m^2$. The result is
\bqa
\label{sc1}
{\cal I}_{\rm sun}^{(1)}(-m^2)&=&{1\over(4\pi)^4}
\left({\mu\over m}\right)^{4\epsilon}
\left[
{1\over\epsilon}J_1T^2+\tilde{K}_1T^2
\right]\;,
\eqa
where $\tilde{K}_1$, which is a function of $\beta m$ only, is defined by
\bqa
\tilde{K}_1=-
{8\over T^2}
\int_0^{\infty}dq\;{q^2n(E_q)\over E_q}
\int_{0}^{1}dx\;\tilde{f}_1(x,q)\;.
\eqa
The function $\tilde{f}_1(x,q)$ in the integrand is 
\bqa
\nonumber 
\tilde{f}_1(x,q) &=& 
{x^2+(1-x)^2\over2x(1-x)q/m}\,
\mbox{atan}\,\left({2x(1-x)q/m\over x^2+(1-x)^2}\right) - 1 \\
&&
+{1\over2}
\log\left(\left[x^2+(1-x)^2\right]^2+4x^2(1-x)^2q^2/m^2\right)\;.
\eqa

\subsubsection{Two thermal factors}

The integral~(\ref{I2-def}) can be expressed as
\bqa
\label{22f2}
{\cal I}_{\rm sun}^{(2)}({\bf p}^2)=
{32\over(4\pi)^4}\int_0^{\infty}dq\;{q^2n(E_q)\over E_q}
\int_0^{\infty}dr\;{r^2n(E_r)\over E_r}
\sum_{\sigma}\mbox{Re}\left\langle
{(-1)\over E_{\sigma}^2-k^2-m^2+i\epsilon}
\right\rangle\;,
\eqa
where $E_{\sigma}=E_q+\sigma E_r$, $k=|{\bf p}+{\bf q}+{\bf r}|$, and
$\sigma$ is summed over $\pm 1$. After performing the angular average, 
(\ref{22f2}) can be analytically continued to ${\bf p}^2=-m^2$. The result is
\bqa
\label{sc2}
{\cal I}_{\rm sun}^{(2)}(-m^2)={1\over(4\pi)^4}\tilde{K}_2T^2\;,
\eqa
where $\tilde{K}_2$, which  is a function of $\beta m$ only, is defined by 
\bqa
\tilde{K}_2&=&{4\over T^2}\int_{0}^{\infty}dq\;{qn(E_q)\over E_q}
\int_{0}^{\infty}dr\;{rn(E_r)\over E_r}
\sum_{\sigma}\tilde{f}_2(E_{\sigma},q,r)\;.
\eqa
The function in the integrand is
\bqa\nonumber
\tilde{f}_2(E,q,r)&=&
\log{[E^2-(q+r)^2]^2+4m^2(q+r)^2\over[E^2-(q-r)^2]^2+4m^2(q-r)^2}
\\ \nonumber
&&
-{2(q+r)\over m}\mbox{atan}{2m(q+r)\over E^2-(q+r)^2} 
+{2|q-r|\over m}\mbox{atan}{2m|q-r|\over E^2-(q-r)^2} \\ 
&&
\label{k2}
+{2\over m}\sqrt{E^2-m^2}\mbox{atan}\;
{8mqr\sqrt{E^2-m^2}\over E^4-2(E^2-2m^2)(q^2+r^2)+(q^2-r^2)^2}\;.
\eqa
If $E^2<m^2$, the last term in (\ref{k2}) should be replaced by a manifestly
real-valued expression using the identity 
$2ix\;\mbox{atan}(ix/y)\longrightarrow x\log[|y-x|/|y+x|]$.

Our final result for the sunset sum-integral evaluated at $p_0=0$ and 
${\bf p}^2=-m^2$ is obtained by combining (\ref{sc0}), (\ref{sc1}), and 
(\ref{sc2}) as in (\ref{sep}):
\bqa\nonumber
{\cal I}_{\rm sun}(0,{\bf p})\Bigg|_{{\bf p}^2=-m^2}&=&
{1\over(4\pi)^4}
\left({\mu\over m}\right)^{4\epsilon}
\Bigg\{
-\left[{3\over2\epsilon^2}+{17\over4\epsilon}-C_1\right]m^2
\\
&&\hspace{2.64cm}
+3\left[
{1\over\epsilon}J_1+\tilde{K}_1+\tilde{K}_2
\right]T^2
\Bigg\}\;.
\eqa
The behavior of the functions $\tilde{K}_1$ and $\tilde{K}_2$ in the limit 
$\beta m\longrightarrow0$ is
\bqa
\tilde{K}_1 &\longrightarrow& {4 \pi^2 \over 3} 
\left[ \log{\beta m \over 4 \pi} + 3 + {\zeta'(-1) \over \zeta(-1)} \right] \;,
\label{Ktilde-1}
\\
\tilde{K}_2&\longrightarrow& -4 \pi^2 
\left[ \log{\beta m \over 4 \pi} - {1 \over 3} + 4 \log 2 
	+ {\zeta'(-1) \over \zeta(-1)} \right] \; .
\label{Ktilde-2}
\eqa 
The result (\ref{Ktilde-1}) was computed analytically.
The result (\ref{Ktilde-2}) was guessed by comparing the expression 
(\ref{twos}) for the screening mass $m_s$ in screened perturbation theory
with the weak-coupling expression for $m_s$ which is given in analytic
form in Ref.~\cite{Braaten-Nieto:scalar}.
It was then verified numerically.

\end{document}